\documentclass[twocolumn,aps,prx,floatfix,superscriptaddress,nofootinbib]{revtex4-2}
\usepackage{amsmath}
\usepackage{amsfonts}
\usepackage{appendix}
\usepackage{graphicx}
\usepackage{dsfont}
\usepackage{amssymb}
\usepackage{bm}
\usepackage{xcolor}
\usepackage{soul}
\usepackage{mathrsfs}
\usepackage{amsmath}
\usepackage{esint}%\usepackage{endfloat}
\usepackage{slashed}
\usepackage{xcolor}
\usepackage{comment}
\usepackage{tikz}

\usepackage{hyperref}
\hypersetup{
    colorlinks=true,
    linkcolor=blue,
    filecolor=Green,      
    urlcolor=blue,
    pdfpagemode=FullScreen,
    citecolor=blue
    }
\usepackage{nccmath}
\usepackage{array}
\usepackage{hyperref}
\newcommand{\beq}{\begin{equation}}
\newcommand{\eneq}{\end{equation}}
\newcommand{\ee}{\end{equation}}
\newcommand{\bea}{\begin{eqnarray}}
\newcommand{\eea}{\end{eqnarray}}

\usepackage{multirow}
\usepackage{wasysym}

\date{\today}

\begin{document}
\title{Dissipation Mechanisms and Dissipative Phase Transitions of two coupled Fully Connected Quantum Ising models}

\author{Bidyut Dey}
\affiliation{I.N.F.N., Gruppo collegato di Cosenza, 
Arcavacata di Rende I-87036, Cosenza, Italy}
\author{Andrea Nava}
\affiliation{Institut f\"ur Theoretische Physik, Heinrich-Heine-Universit\"at, 40225 D\"usseldorf, Germany}
\author{Domenico Giuliano}
\affiliation{I.N.F.N., Gruppo collegato di Cosenza, 
Arcavacata di Rende I-87036, Cosenza, Italy}
\affiliation{Institut f\"ur Theoretische Physik, Heinrich-Heine-Universit\"at, 40225 D\"usseldorf, Germany}
\affiliation{Dipartimento di Fisica, Universit\`a della Calabria Arcavacata di 
Rende I-87036, Cosenza, Italy}
 
\begin{abstract}
We study dissipative phase transitions in a system of two coupled fully-connected quantum Ising models interacting with an environment. The dynamics is governed by a Lindblad master equation combining coherent unitary evolution and incoherent dissipative processes, where the unitary part is described within a self-consistent mean-field framework  effectively acting on the local Hilbert space of two coupled spins at each site. 

We analyze two fundamentally different classes of dissipators. In the first case, the jump operators are defined in the instantaneous eigenbasis of the mean-field Hamiltonian and satisfy a detailed-balance condition. In this setting, the relaxation dynamics depends strongly on the quench protocol: a parametric quench of the Hamiltonian leads to conventional relaxation, whereas a temperature quench gives rise to a dynamical phase transition characterized by nonanalytic behavior in time. Yet, in both cases, the system relaxes toward a steady state determined solely by the post-quench parameters and the bath temperature, which closely resembles a thermal Gibbs state of the mean-field Hamiltonian. As a result, the dissipative phase transition occurs at a critical point consistent with the corresponding equilibrium transition. In contrast, when the dissipators are realized via local spin raising and lowering operators, the steady state is genuinely nonequilibrium, leading to a significantly richer phase diagram. In particular, for sufficiently strong system–bath coupling, we observe a reentrant phase featuring a symmetry-broken region bounded by two continuous dissipative phase transitions.

Our results evidence how the structure of dissipative processes controls the emergence of equilibrium-like versus genuinely nonequilibrium critical behavior in open quantum systems.  

\end{abstract}

\date{\today}

\maketitle

\section{Introduction}

 When a quantum system is coupled to an external environment, the competition between coherent unitary dynamics and incoherent dissipative processes can drive the system into nontrivial, equilibrium or nonequlibrium,  steady states, whose specific properties are determined by the system Hamiltonian, by  the nature of the environment and by the coupling between the environment and the system itself \cite{Guimaraes2016,Pizorn2013,Benenti2009,Nava2021,Cinnirella2024,Cinnirella2025,Cinnirella2026}. Remarkably, these  nonequilibrium steady states (NESS)s, have no direct analog in equilibrium statistical mechanics and can exhibit rich collective behavior that is absent in isolated systems. In particular,  the value of observables computed in the NESS can display a nonanalytic behavior as system parameters are varied, signaling the occurrence of a dissipative phase transition (DPT), determined by the coupling between the system and the environment.

Recently, DPTs in open quantum systems have attracted considerable attention. In particular, they have been extensively studied in a variety of platforms, including driven photonic systems\cite{PhysRevX.5.031028, PhysRevA.96.023839, PhysRevA.94.033841, Kewming_2022}, interacting spin systems \cite{PhysRevA.86.012116, PhysRevX.6.031011, current, Optical_Cavity_QED, Kothe:2025, Joshi_2013, Kshetrimayum_2017} and have also been observed experimentally \cite{Fink:2016}. These transitions can be either first-order or second-order \cite{Entropy_Production_2012, Entropy_Production_2026, Beaulieu_2025, Tim_2018}. Moreover, nonequilibrium phase transitions in open quantum systems have also been identified in boundary-driven and periodically driven spin chains \cite{prosen2008, prosen2011}, where criticality manifests through the emergence of long-range correlations and changes in entanglement properties of the steady state. Normally, the dynamics of an open quantum system coupled to a memoryless (Markovian) environment is typically described by the Lindblad master equation (LME) \cite{Lindblad1976,Breuer2007}. In isolated quantum systems, phase transitions are identified through nonanalyticities in the spectrum of the Hamiltonian, which governs the unitary time evolution. By contrast, in open quantum systems the relevant generator of dynamics is the Liouvillian superoperator, which governs the time evolution of the density matrix $\rho$ and generally possesses a complex spectrum \cite{Prosen:2008}. A central question in the study of dissipative phase transitions concerns the identification of quantities that universally characterize critical behavior. In linear open quantum systems \cite{Nigro:2018, Victor_2014}, the spectral properties of the Liouvillian provide such a diagnostic. In particular, Ref.~\cite{PhysRevA.98.042118} introduced the Liouvillian gap, defined as the real part of the smallest nonzero eigenvalue of the Liouvillian superoperator. The closing of this gap signals critical slowing down and can be used to distinguish between first-order (discontinuous) and second-order (continuous) dissipative phase transitions \cite{PhysRevA.97.013853, PhysRevB.101.214302, PhysRevA.86.012116}. 

However, the Liouvillian gap alone is not always sufficient to characterize DPTs \cite{Taiki_Haga}, especially in situations where the  Liouvillian depends self-consistently on the $\rho$ and, therefore, the whole set of equations becomes intrinsically nonlinear \cite{Sarandy:2005}. In such cases, the spectral properties of the Liouvillian no longer provide a complete description of the dynamics. One must instead rely on steady-state observables and information-theoretic quantities to probe critical behavior such as, for instance, the fidelity susceptibility \cite{PhysRevA.105.052226}, the quantum Fisher information \cite{Fernandez_2017}, DMRG \cite{Diehl_2016} and  entanglement measures \cite{PhysRevB.95.134431, Wim_2017}, as well as approaches based on cluster mean-field methods \cite{PhysRevX.6.031011, PhysRevB.104.155130, PhysRevB.97.035103}. These quantities provide complementary signatures of dissipative criticality and allow one to identify phase transitions even in nonlinear driven-dissipative systems. 

Remarkably, it has been shown that dissipation can be engineered as a resource to prepare and stabilize complex many-body quantum states and phases \cite{Verstraete:2008, Diehl:2008}. In particular, appropriately designed system–bath couplings can drive the system into desired steady states, including strongly correlated and long-range ordered phases, thereby providing a route to realizing nonequilibrium quantum phases through reservoir engineering \cite{Muller:2012}. In this respect, a key question in open quantum systems is how the nature of dissipation influences the steady-state properties and the resulting phase transitions. In particular, it is important to understand under what conditions dissipative dynamics reproduces equilibrium-like behavior and when it leads to genuinely nonequilibrium phenomena. This distinction depends sensitively on the structure of the dissipative channels and the resulting interplay between coherent and incoherent processes. 

In this work, we address the above issue by comparing two distinct classes of dissipative mechanisms, revealing how the structure of dissipation determines whether the system exhibits equilibrium-like or genuinely nonequilibrium critical behavior. Specifically, we consider a system of two coupled, fully connected one-dimensional quantum Ising models (QIMs). In the thermodynamic limit, the mean-field description of the isolated system becomes exact, and the many-body density matrix factorizes into identical single-site density matrices acting on the local Hilbert space of the two spins at each site, one from each model. The resulting mean-field Hamiltonian depends self-consistently on the density matrix, thereby rendering the dynamics intrinsically nonlinear \cite{Fabrizio2018,Nava2022}.  

We first consider dissipative dynamics generated by jump operators defined in the instantaneous eigenbasis of the mean-field Hamiltonian, supplemented by a detailed balance condition. Detailed balance ensures that transitions between eigenstates satisfy thermal constraints, such that the steady state takes a Boltzmann form in the eigenbasis of the mean-field Hamiltonian. Since the self-consistent mean-field (SCMF) equations become time dependent out of equilibrium, the resulting Lindblad master equation (LME) is both nonlinear and explicitly time dependent. Although the steady state is thermal in form, this behavior, as evidenced in several studies employing time-dependent SCMF approaches, arises from the underlying nonlinear, self-consistent dissipative dynamics~\cite{Nava2023_s,Nava2024_s,Nava2025_s,Giuliano2026}.

 Within this framework, we investigate the relaxation dynamics following both a parametric quench and a temperature quench in the presence of dissipation. In particular, we examine whether these protocols drive the system through dynamical phase transitions, characterized by nonanalytic behavior in the time evolution of observables and in the fidelity rate function. We first consider a quench of the transverse field in the mean-field Hamiltonian while keeping all other parameters and the bath temperature fixed. In this case, the post-quench evolution exhibits conventional relaxation characterized by a smooth crossover. In contrast, a bath temperature quench, performed while keeping the Hamiltonian parameters fixed, leads to qualitatively different dynamics. For weak dissipation, cusp-like nonanalyticities emerge in the time evolution of observables and in the fidelity rate function, signaling a dynamical phase transition. As the dissipation strength increases, these nonanalytic features are progressively smoothed out. 

 The dissipative dynamics is governed by the LME, which we numerically integrate to obtain the asymptotic steady state. Due to the detailed-balance structure of the jump operators, the steady state coincides with a thermal Gibbs state associated with the mean-field Hamiltonian, and all the observables are computed directly from its density matrix. At low bath temperatures, the system is driven toward a state closely related to the equilibrium mean-field ground state. As a result, the critical point of the dissipative phase transition coincides with the equilibrium quantum phase transition point.

We analyze the properties of the steady state obtained via self-consistent dissipation and identify the phase transition through nonanalytic behavior of observable quantities as functions of the tuning parameters. In particular, the order parameter vanishes continuously at the critical point, signaling a continuous phase transition \cite{Pedro_LMG, Optical_Cavity_QED, Kothe:2025}. In the presence of nonlinear Lindblad dynamics, the Liouvillian becomes state dependent, and the stability gap obtained from the Jacobian provides the relevant generalization. The transition is therefore further characterized via linear stability analysis, where the Jacobian evaluated at the steady state determines its stability and the associated relaxation time scales \cite{Debecker_2025, Debecker_2024, Kothe:2024}. We find that the stability gap remains finite and positive across the entire range of the tuning parameter, indicating the absence of true critical slowing down. Nevertheless, near the critical point we observe a pronounced but finite slowing down of the relaxation dynamics.

We then consider an alternative dissipative mechanism using local spin raising and lowering operators as jump operators. In this case, the dissipators are no longer defined self-consistently in the eigenbasis of the mean-field Hamiltonian and therefore do not satisfy detailed balance. As a result, the steady state is genuinely a non-equilibrium steady state (NESS) and cannot be associated with a thermal state. For this case, instead of studying the full density matrix dynamics, we focus on   properties of physically relevant observables computed in the NESS   \cite{Optical_Cavity_QED, Optical_Cavity_QED_2008}. By deriving a closed set of Bloch-type equations for local magnetizations and correlation functions, we obtain the steady-state solutions directly without integrating the LME. The linear stability analysis is then performed within this reduced observable space by evaluating the Jacobian of the coupled equations \cite{Francois_2019, Kothe:2024}. This approach captures the collective behavior of the system and allows us to identify DPTs through the stability of steady-state observables. The competition between coherent dynamics and dissipation results in a richer phase diagram. While at weak system–bath coupling the critical point coincides with the equilibrium one, increasing the dissipation strength leads to a reentrant phase diagram, highlighting the nontrivial role of dissipation in stabilizing and destabilizing ordered phases.

The two approaches thus correspond to fundamentally different dissipative mechanisms: self-consistent thermalization vs local pump-loss dynamics. Our results demonstrate how the nature of dissipation qualitatively alters the structure of dissipative phase transitions in mean-field quantum systems. Although our analysis is based on a mean-field framework, it captures generic features of driven-dissipative systems, and similar behavior is expected in systems with long-range interactions.

The paper is organized as follows:

\begin{itemize}

\item In Section \ref{modam} we present our model Hamiltonian for the coupled fully connected quantum Ising model and review its basic properties, including the phase diagram of the isolated system.

\item In Section \ref{lima} we review LME approach to the dissipative dynamics of the open system.

\item In Section \ref{r_ge_1} we discuss the post-quench relaxation dynamics of the system with jump operators defined in the instantaneous eigenbasis of the mean-field Hamiltonian.

\item In Section \ref{sec:DPT_A} we study the dissipative phase diagram of the system with jump operators defined as above.

\item In Section \ref{DPT_B} we discuss the dissipative phase diagram of the system with local dissipators.

\item In Section \ref{concl} we provide our conclusions and discuss about further perspectives of
our work.

\end{itemize}

\section{Model Hamiltonian}
\label{modam}

Throughout this paper, we discuss the model Hamiltonian describing two, fully connected, $N$-site one-dimensional quantum Ising models (QIMs), coupled to each other by means of a local spin exchange interaction. The model Hamiltonian is accordingly  given by \cite{Fabrizio2018,Nava2022}
\begin{equation}
H = \sum_{n=1}^{2} \hat{H}_n - \lambda \sum_{j=1}^{N} \sigma^{x}_{1,j} \sigma^{x}_{2,j}
\;\: ,
\label{model_H}
\end{equation}
\noindent
with the Hamiltonian of each one-dimensional QIM ($n=1,2$) given by 
\begin{equation}
\hat{H}_n = -\frac{J_n}{2N} \sum_{i,j=1}^{N} \sigma^{x}_{n,i} \sigma^{x}_{n,j} - h_n \sum_{i=1}^{N} \sigma^{z}_{n,i}
\label{local_H}
\:\:.
\end{equation}
\noindent
In Eqs.(\ref{model_H},\ref{local_H}) $\sigma^{a}_{n,i}$ ($a=x,y,z$, $n=1,2$, $i=1,\ldots ,N$) denote the Pauli matrices acting on site $i$ of model~$n$. All the system parameters, namely, the coupling strengths $J_n$, the transverse fields $h_n$, and the inter-model coupling $\lambda$ are taken to be positive and, throughout this work we assume $J_1=J_2$ and set $J_1 = J_2 = 1$ to fix our energy scale.

In reviewing the equilibrium phase diagram of the isolated system we note that the full connection of the system corresponds to an infinite-range interaction. In this limit, one finds that  the connected two-spin correlation functions scale as the inverse power of the system size, that is 
\begin{equation}
\langle \sigma^{a}_{n,i} \sigma^{b}_{n,j} \rangle - \langle \sigma^{a}_{n,i} \rangle \langle \sigma^{b}_{n,j} \rangle \sim \frac{1}{N}, \qquad i\neq j,
\label{scaling_corr}
\end{equation}
\noindent
and therefore, they vanish in the thermodynamic limit $N\to\infty$ and, accordingly, the mean-field theory becomes exact \cite{Das2006,Bapst2012,Fabrizio2018}. Thus, as $N\to \infty$, the density matrix $\hat{\rho}$ of the system factorizes as
\begin{equation}
\hat{\rho} \xrightarrow[N\to\infty]{} \bigotimes_{i=1}^{N} \hat{\rho}_i 
\label{defac}
\;\; ,
\end{equation}
\noindent
with each $\hat{\rho}_i$ acting on the combined Hilbert space of site $i$ in model~1 and site~$i$ in model~2 and, thus, being a positive-definite $4\times4$ matrix with unit trace. That being stated, we now discuss the phase diagram of the system, starting from the disconnected limit ($\lambda = 0$) and then extending our analysis to the case $\lambda > 0$. 

\subsection{Equilibrium phase diagram in the disconnected  limit}
\label{eqdisc}

\begin{figure}
\centering
\includegraphics[width=0.75\linewidth]{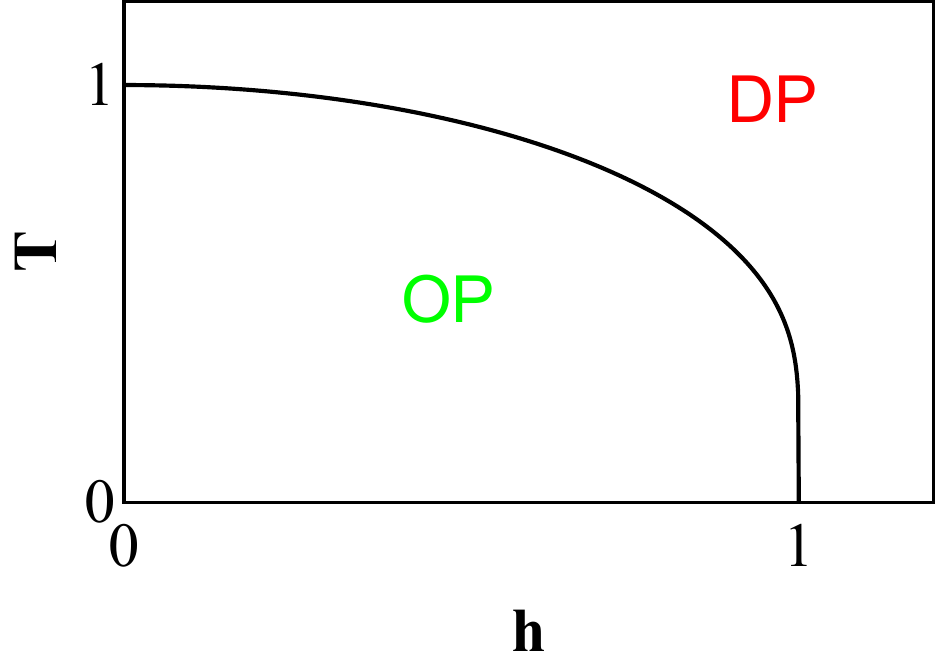}
\caption{Phase diagram of the fully connected quantum Ising model in Eq.~\eqref{1Ising_Ham}. 
The black curve denotes the critical temperature given by Eq.~\eqref{1Ising_Tc}. Below this line the system is in the ordered phase (OP) with spontaneous $\mathcal{Z}_2$ symmetry breaking, while above it the system is in the disordered phase (DP).}
\label{fig:1Ising_phase_diagram}
\end{figure}

As $\lambda = 0$, we may readily derive the phase diagram within mean field approximation, by taking advantage of the density matrix factorization in Eq.(\ref{defac}). Clearly, as now the two one-dimensional QIMs are disconnected from one another, we may just analyze the phase diagram of a single fully connected model. Accordingly, in this Section we focus on the Hamiltonian $\hat{H}$, given by 

\begin{equation}
\hat{H} = -\frac{1}{2N} \sum_{i,j=1}^{N} \sigma^{x}_{i} \sigma^{x}_{j} - h \sum_{i=1}^{N} \sigma^{z}_{i}
\:\:.
\label{1Ising_Ham}
\end{equation}
\noindent
Consistently with the approach developed in Refs.\cite{Fabrizio2018,Nava2022}, we neglect fluctuations in the value of the $\sigma_i^x$ and, therefore, up to an over-all, unessential constant, we  approximate $\hat{H} \approx \sum_{i=1}^N \hat{H}_i$, with 
\begin{equation}
\hat{H}_i = - m_x (T) \sigma_i^x - h \sigma_i^z 
\;\: , 
\label{single_siteh}
\end{equation}
\noindent
and $m_x (T) = \langle \sigma_i^x \rangle$. The single-site mean-field Hamiltonian $\hat{H}_i$ in Eq.(\ref{single_siteh}) describes an effective two-level system at each site. Its two eigenstates are separated by an energy gap
\begin{equation}
E(T) = 2\,h_{\mathrm{eff}} = 2\,\sqrt{m_x^2(T) + h^2} \:\: . 
\label{eq:MF_gap}
\end{equation}
\noindent
Since the infinite-range interaction suppresses spatial correlations in the thermodynamic limit, the resulting excitation spectrum is dispersionless. The gap, therefore, corresponds to an optical-like collective excitation rather than a propagating quasiparticle mode. Using $\hat{H}_i$, we compute the magnetization $m_x(T)$, which serves as the order parameter for the magnetic phase transition in the quantum spin model. This leads to the following self-consistent equation of state:
\begin{equation}
m_x (T) = \frac{m_x (T) \tanh [ \beta \sqrt{(m_x (T))^2 + h^2} ]}{ \sqrt{(m_x (T))^2 + h^2}}
\:\: , 
\label{stateq}
\end{equation}
\noindent
with $\beta = (k_B T)^{-1}$ and $k_B$ being Boltzmann constant. From  Eq.(\ref{stateq}) one recovers the equilibrium phase diagram shown in Fig.~\ref{fig:1Ising_phase_diagram}. For $h\le1$, the critical temperature is
\begin{equation}
k_B T_c = \frac{2h}{\ln\!\left(\frac{1+h}{1-h}\right)}
\:\:.
\label{1Ising_Tc}
\end{equation}
\noindent
For $h\le1$ and $T < T_c$, a nonzero magnetization $m_x$ spontaneously breaks the $\mathcal{Z}_2$ symmetry $\sigma_i^x \rightarrow -\sigma_i^x$, defining the ordered phase. For $h\ge1$ or $T > T_c$, the symmetry is restored and the system is disordered. In the ordered phase ($h<1$ and $T<T_c$), the system exhibits a finite spontaneous magnetization $m_x(T)$. At zero temperature one has
\begin{equation}
m_x(0) = \sqrt{1-h^2}
\;\;,
\label{teq0}
\end{equation}
\noindent
which yields a zero-temperature excitation gap
\begin{equation}
E(0) = 2\,\sqrt{m_x^2(0) + h^2} = 2
\:\:.
\label{0tex}
\end{equation}
\noindent
Thus, at $T=0$ and for $h<1$, the excitation gap is dominated by the self consistently generated collective mean field and is pinned to a constant value, independent of the transverse field. At finite temperature, thermal fluctuations suppress the spontaneous magnetization, leading to a reduction of the excitation gap with increasing temperature. At the critical temperature $T_c(h)$, the magnetization vanishes while the gap remains finite, taking the paramagnetic value $E=2h$. For temperatures above the critical line, $T>T_c(h)$, the system enters the disordered phase characterized by $m_x(T)=0$. In this regime the excitation gap becomes temperature independent and is solely determined by the transverse field,
\begin{equation}
E(T) = 2h \:\:.
\label{solely}
\end{equation}
\noindent
Physically, this behavior reflects the fact that in the ordered phase the dominant energy scale is set by the self-consistent mean field generated through spontaneous symmetry breaking, whereas in the disordered phase the dynamics is governed entirely by the external transverse field. The absence of momentum dependence highlights the collective, non-propagating nature of the elementary excitations in the fully connected quantum Ising model. 
   
\noindent   
 We now review how the phase diagram discussed in this Section is affected by turning on $\lambda$. 
 
 \subsection{Equilibrium phase diagram at finite inter-model coupling}
 \label{finter}

For $\lambda > 0$, the effective single-site Hamiltonian $\hat{H}_i$ is replaced by a temperature-dependent operator acting on the local Hilbert space of the two coupled models, $\hat{H}_{\mathrm{MF},i}(T)$. The corresponding single-site equilibrium density matrix is defined as
\begin{eqnarray}
\hat{H}_{\mathrm{MF},i}(T) &=&  -\sum_{n=1}^2 \left[ 
J_n^{\mathrm{eq}}(T)\,\sigma^x_{n,i} + h_n \sigma^z_{n,i} \right] 
- \lambda \,\sigma^x_{1,i}\sigma^x_{2,i} \:\: ,
\nonumber \\
\hat{\rho}_i^{\mathrm{eq}}(T) &=& 
\frac{e^{-\beta \hat{H}_{\mathrm{MF},i}(T)}} 
{\mathrm{Tr}\!\left[e^{-\beta \hat{H}_{\mathrm{MF},i}(T)}\right]}
\:\: ,
\label{1Ising_SC}
\end{eqnarray}
with the mean-field couplings determined self-consistently as
\begin{equation}
J_n^{\mathrm{eq}}(T) = \lim_{N\to\infty}\frac{1}{N}\sum_{j=1}^{N} 
\mathrm{Tr}\!\left(
\hat{\rho}_j^{\mathrm{eq}}(T)\sigma^x_{n,j} \right) 
\equiv m_{x,n}(T)
\:\: ,
\label{mfcouplings}
\end{equation}
which is site-independent, due to translational invariance.

Following the analysis of Ref.~\cite{Nava2022}, we focus on the parameter regime $\lambda \ll h_1 \ll h_2$. When $h_1 < 1$ and $T < T_c(h_1)$, where $T_c(h_1)$ is the critical temperature of the QIM~1, the system develops a spontaneous magnetization $m_{x,1}(T) \neq 0$. A finite $\lambda > 0$ induces a small magnetization in the QIM~2, given by
\begin{equation}
m_{x,2}(T) \simeq \frac{\lambda}{h_2}\, m_{x,1}(T)
\:\: .
\label{finmat}
\end{equation}
Clearly, although being finite, $m_{x,2} (T)$ remains parametrically suppressed by the 
condition  $h_2 \gg h_1$, that is, the large value of $h_2$ suppresses the feedback from the second model. As a result, the equilibrium phase diagram remains approximately unchanged, as compared to the $\lambda = 0$ limit. 

To leading order in $\lambda$, the local mean-field Hamiltonian therefore supports two dispersionless excitation branches with energies $E_{1,2} (T)$ given by 
\begin{eqnarray}
E_1(T) &\simeq& 2\sqrt{m_{x,1}^2(T)+h_1^2} \:\: , \nonumber\\
E_2(T) &\simeq& 2h_2 \gg E_1(T)
\:\:.
\label{dispersionless}
\end{eqnarray}
\noindent
$E_1 (T)$ and $E_2 (T)$ correspond to optical, non-propagating excitations associated with collective spin flips in models~1 and~2, respectively. $\hat{H}_{\rm MF,i}$ can be conveniently represented in terms of two species of hard-core bosons, 
$b_{n,i}$ ($n=1,2$), such that  
\begin{equation}
\hat{H}_{\rm MF,i} = \sum_{i=1}^{N}\sum_{n=1}^{2} E_n(T)\, b^\dagger_{n,i} b_{n,i}
\:\:,
\label{mfhi}
\end{equation}
\noindent
with  eigenstates
\begin{equation}
|k,i\rangle = \left(b^\dagger_{1,i}\right)^{n_1(k)} \left(b^\dagger_{2,i}\right)^{n_2(k)} |0\rangle, \qquad k=0,1,2,3
\;\;,
\label{eigenstates}
\end{equation}
\noindent
and corresponding energies
\begin{equation}
E(k) = n_1(k) E_1(T) + n_2(k) E_2(T)
\;\;,
\label{energies}
\end{equation}
\noindent
with  $n_2(k)=\lfloor k/2 \rfloor$ and $n_1(k)=k-2n_2(k)$.
The low-energy sector, spanned by the states $|0,i\rangle$ and $|1,i\rangle$, is well separated from the high-energy sector containing $|2,i\rangle$ and $|3,i\rangle$ by the large gap $E_2(T)\simeq 2h_2$. Consequently, there exists a wide temperature window
\begin{equation}
T_c(h_1) \lesssim T \ll E_2(T)=2h_2,
\end{equation}
in which the thermodynamics is dominated by the low-energy sector, which carries a finite entropy, while the high-energy sector remains essentially unpopulated. This separation of energy scales provides a natural setting for introducing controlled local dissipation, as we discuss in the following. 

\section{Lindblad approach to the dissipative dynamics}
\label{lima}

To initiate the relaxation dynamics of the system coupled to an external reservoir, we first prepare the system in a specific equilibrium state at $t=0^-$. At $t=0$, we perform a parametric quench on the system Hamiltonian, which induces intrinsically nonequilibrium dynamics, and turn on the coupling between the system and the external reservoir. This coupling drives the system toward an asymptotic steady state whose nature strictly depends on the details of the dissipation mechanism (the ``dissipators"). Due to the intrinsically nonequilibrium relaxation dynamics, the density matrix operator $\hat{\rho}_i(t)$ in Eq.(\ref{1Ising_SC}) becomes explicitly time-dependent. We determine its time evolution using Lindblad master equation (LME) approach \cite{Kosov2011,Lang2015}.  Specifically, following the method developed in \cite{Peronaci2015,Nava2023_s,Nava2024_s}, we employ a nonlinear, explicitly time-dependent LME for $\hat{\rho}_i(t)$ given by 

\begin{equation}
\frac{\partial \hat{\rho}_i}{\partial t} = -i\bigl[ \hat{H}_{\rm{MF},i} (\hat{\rho}_i), \hat{\rho}_i \bigr] + \sum_{n<m} \left( \mathcal{D}_{n\leftarrow m}[\hat{\rho}_i] + \mathcal{D}_{n\rightarrow m}[\hat{\rho}_i] \right)\;\; . 
 \label{eq:nonlinear_lindblad}
\end{equation}
\noindent
As stated above, in Eq.(\ref{eq:nonlinear_lindblad}), the Hamiltonian depends self-consistently on the density matrix through the time-dependent order parameters,
\begin{equation}
\hat{H}_{\mathrm{MF},i}(\hat{\rho}_i) = -\sum_{n=1}^{2} \left[ m_{x,n}(t)\,\sigma^x_{n,i} + h_n\,\sigma^z_{n,i} \right] -\lambda\,\sigma^x_{1,i}\sigma^x_{2,i}\;\;, 
\label{eq:Main_MF_H}
\end{equation}
\noindent
with the magnetization $m_{x,n} (t)$ determined according to  
\begin{equation}
m_{x,n}(t) = \mathrm{Tr}\!\left( \hat{\rho}_i(t)\,\sigma^x_{n,i} \right)\:\:. 
\label{eq:m1}
\end{equation}
\noindent
Here, we construct the dissipative contribution on the right-hand side of Eq. (\ref{eq:nonlinear_lindblad}) (the 'Lindbladian') to drive the system toward thermal equilibrium. In the following, we discuss the effect of an alternative definition of the `` jump operators''. Specifically, we set 
\begin{eqnarray}
\mathcal{D}_{n\leftarrow m}[\hat{\rho}_i] &=& \gamma_{n\leftarrow m}(t) \big( \hat{L}_{n\leftarrow m}(t)\hat{\rho}_i(t)\hat{L}_{n\rightarrow m}(t) \nonumber\\&-&\frac{1}{2} \bigl\{ \hat{L}_{n\rightarrow m}(t)\hat{L}_{n\leftarrow m}(t), \hat{\rho}_i(t) \bigr\} \big) \nonumber \:\: , \\
\mathcal{D}_{n\rightarrow m}[\hat{\rho}_i] &=& \gamma_{n\rightarrow m}(t) \big( \hat{L}_{n\rightarrow m}(t)\hat{\rho}_i(t)\hat{L}_{n\leftarrow m}(t) \nonumber\\&-&\frac{1}{2} \bigl\{ \hat{L}_{n\leftarrow m}(t)\hat{L}_{n\rightarrow m}(t), \hat{\rho}_i(t) \bigr\} \big)\;\; ,
\label{lindbladian}
\end{eqnarray}
\noindent
with the jump operators defined in the instantaneous eigenbasis of $\hat{H}_{\mathrm{MF},i}(\hat{\rho}_i)$, $\{|n;i,t\rangle\}$ ($n=0,1,2,3$), with eigenvalues $\{E_n (t)\}$ ordered at $t=0$ such that $E_0 (0) \le E_1 (0) \le E_2 (0) \le E_3 (0)$, and given by
\begin{equation}
\hat{L}_{n\leftarrow m}(t) = |n;i,t\rangle\langle m;i,t|, \qquad n<m
\:\: . 
\label{jump_eigen}
\end{equation}
\noindent
 For $n<m$, the operator $\hat{L}_{n\leftarrow m}$ in Eq.(\ref{jump_eigen})  describes a downward transition (de-excitation) from a higher- energy state to a lower-energy one, while the reverse process is generated by $\hat{L}_{n\rightarrow m}=\hat{L}_{n\leftarrow m}^\dagger$. Finally, the transition rates are defined to satisfy the detailed balance condition
\begin{equation}
\frac{\gamma_{n\rightarrow m}(t)} {\gamma_{n\leftarrow m}(t)} = e^{-\beta\left[E_m(t) -E_n(t)\right]}
\:\:, \label{eq:detailed_balance}
\end{equation}
\noindent
which guarantees that, in the absence of mean-field feedback, the dissipative dynamics relaxes the system at late times toward the Gibbs state associated with $\hat{H}_{\mathrm{MF},i}$.

Eq.(\ref{eq:nonlinear_lindblad}) is solved with the initial condition $\hat{\rho}_i(0)=\hat{\rho}_i^{\mathrm{eq}}(T)$, which is identical for all sites $i$. As a consequence of translational invariance (built in the self-consistent approximation), the density matrix remains site independent at all later times. The Lindblad dynamics is fully specified by the six downward transition rates $\gamma_{n\leftarrow m}$ with $n<m$, among which we distinguish between high-energy and low-energy relaxation processes by assuming
\begin{equation}
\gamma_{n\leftarrow m} = \begin{cases} \gamma_{\mathrm{high}}^{\Downarrow},
 & (n,m)\in\{(0,2),(0,3),(1,2),(1,3)\},\\[4pt] \gamma_{\mathrm{low}}^{\Downarrow}, 
 & (n,m)\in\{(0,1),(2,3)\}\:\:, 
 \end{cases}
 \label{rates}
\end{equation}
\noindent
and
\begin{equation}
\frac{\gamma_{\mathrm{high}}^{\Downarrow}}{\gamma_{\mathrm{low}}^{\Downarrow}} = r \ge 1\;\;,
\label{assumption_r}
\end{equation}
\noindent
 which reflects the assumption that relaxation processes involving the high-energy sector occur on shorter time scales than those within the low-energy sector. As a result, population in the high-energy states rapidly decays toward the low-energy manifold, leading to an effective cooling-like dynamics. The LME in Eq.(\ref{eq:nonlinear_lindblad}) is valid in the regime of weak system--bath coupling and finite temperature, which we assume to hold in the case we are considering here. 
  
As a general comment, it is worth mentioning that, while here we focused on a specific regime of values of the system parameters, in which the features we are interested in become particularly evident, there is also the possibility to address the alternative parameter regime in which  $h_2 \sim h_1$ and $\lambda \sim h_1$. In this case, the local spectrum does not exhibit a clear separation of energy scales, and all four energy levels become comparable, that is 
\begin{equation}
E_0 \sim E_1 \sim E_2 \sim E_3
\:\: .
\label{comparable_energies}
\end{equation}
\noindent
When the conditions in Eq.(\ref{comparable_energies}) hold, there is no well-separated low-energy sector and the entropy is distributed across the full local Hilbert space. Consequently, the relaxation rates satisfy $\gamma_{\mathrm{high}}^{\Downarrow}\sim\gamma_{\mathrm{low}}^{\Downarrow}$, while the detailed balance condition~\eqref{eq:detailed_balance} remains valid and continues to guarantee relaxation toward the Gibbs state. 

Having established the equilibrium phase diagram and the LME description of the dissipative dynamics we now study the post-quench relaxation dynamics across the phase transition line. 

\section{Post-quench relaxation dynamics}
\label{r_ge_1}
 
 We now investigate the post-quench dynamics of our system as determined by the LME in Eq. \eqref{eq:nonlinear_lindblad}, supplemented by the time-dependent SCMF Hamiltonian in Eqs. \eqref{eq:Main_MF_H} and \eqref{eq:m1}. In particular, we focus on two main quench protocols. First, we consider a quench in the transverse field of model~1, $h_1$, while keeping the bath temperature fixed. In this case, the post-quench time evolution exhibits standard relaxation behavior without signatures of dynamical phase transitions. Second, we consider a sudden change in the bath temperature, while keeping the Hamiltonian  parameters fixed. In this case, the post-quench dynamics triggers a nonanalytic change at a critical time, signaling a dynamical phase transition. Similar dynamical phase transitions in driven-dissipative systems have been reported in~\cite{Nava2023_s,Nava2024_s,Wu2024, Nava2025_s}.

As stated above, we focus on the regime $h_2 \gg h_1$ with a weak inter-model coupling $\lambda \ll 1$. In this limit, the local Hilbert space at each site separates into two well-separated energy sectors. This hierarchy of energy scales allows us to treat the low-energy sector independently. Moreover, the dissipative rates are engineered such that transitions involving the high-energy sector occur at larger rates than those within the low-energy sector. Such a separation of energy scales, combined with the controlled choice of dissipative channels, provides a natural setting for introducing controlled dissipation. To characterize the relaxation dynamics, we compute the time-dependent post-quench local order parameter $m_{x,1}(t)$, which provides information on the single-model state, and the inter-model correlator $O(t) = \langle \sigma^x_1 \sigma^x_2\rangle$, which allows for mapping out how inter-model correlations evolve in time. Throughout our preparation protocol, we fix $h_2$ to a large value, $h_2 = 10$, and choose a weak inter-model coupling, $\lambda = 0.01$, such that the assumptions leading to Eq. \eqref{assumption_r} are satisfied. Under these conditions, relaxation processes involving large energy differences occur on much shorter time scales than those within the low-energy sector. Consequently, the equilibrium phase diagram in Fig. \eqref{fig:1Ising_phase_diagram} remains approximately unchanged.  At $t=0$, we prepare the system in an ordered thermal equilibrium state by choosing the initial values of the transverse field $h_{1,i}$ and temperature $T_i$ such that the state lies in the ordered phase, characterized by a finite magnetization $m_{x,1}$ (see Appendix~\ref{Initial_thermal} for details). Owing to the finite $\lambda$, this ordering in the low-energy sector induces a small but finite magnetization $m_{x,2}$ in model~2. We then perform a parametric quench by suddenly changing the system parameters and evolving the density matrix under the LME corresponding to the post-quench Hamiltonian, and the dissipative channels. The quench is chosen to drive the system across the equilibrium phase boundary into the disordered regime. Since the initial state is not stationary under the post-quench dynamics, the system undergoes nonequilibrium relaxation toward a new steady state \cite{Nava_2023, Giuliano2026}. The subsequent time evolution of the density matrix is obtained by solving the self-consistent mean-field Lindblad equations, Eqs.~\eqref{eq:nonlinear_lindblad}, \eqref{eq:Main_MF_H}, and \eqref{eq:m1}. Also, consistently with the expected validity of the LME in the low system-bath coupling, we choose $\gamma^{\Downarrow}_{\rm{high}} =0.05$. As a first result of our calculation, we recover the post-quench relaxation dynamics of $m_{x,1}(t)$.

\begin{figure}[t]
\centering

\begin{tikzpicture}
\node (A) at (0,0) {
\begin{tabular}{@{}cc@{}}
\includegraphics[width=0.46\linewidth]{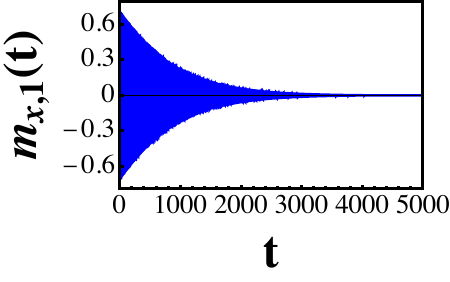} &
\includegraphics[width=0.46\linewidth]{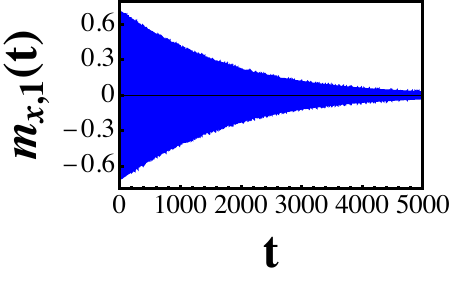}
\end{tabular}
};
\node at (A.north west) [anchor=north west, xshift=-4pt, yshift=2pt] {\textbf{(a)}};
\end{tikzpicture}

\vspace{0.3cm}

\begin{tikzpicture}
\node (B) at (0,0) {
\begin{tabular}{@{}cc@{}}
\includegraphics[width=0.46\linewidth]{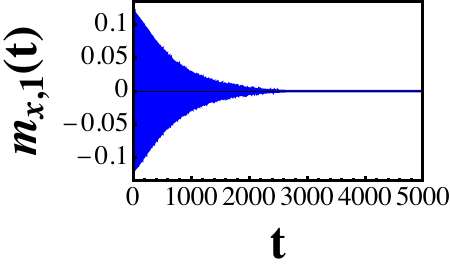} &
\includegraphics[width=0.46\linewidth]{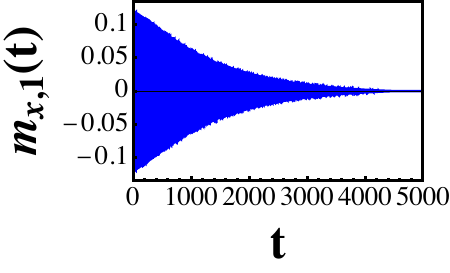}
\end{tabular}
};
\node at (B.north west) [anchor=north west, xshift=-4pt, yshift=2pt] {\textbf{(b)}};
\end{tikzpicture}

\caption{Post-quench time dependence of the order parameter $m_{x,1}(t)$ for 
$r=10$ (left) and $r=20$ (right). At $t=0$, a sudden quench of the transverse 
field is performed from $h_{1,i}=0.7$ to $h_{1,f}=1.1$, while the bath 
temperature is kept fixed at (a) $T=0.05$ and (b) $T=0.8$. All parameters are 
given in units of $J=1$.}
\label{fig:1_h1_quench}

\end{figure}
Figure~\ref{fig:1_h1_quench} shows the time evolution of $m_{x,1}(t)$ following a sudden quench of the transverse field in model~1, $h_{1,i} = 0.7 \to h_{1,f} = 1.1$, for fixed bath temperatures $T = 0.05$ [Fig.~\ref{fig:1_h1_quench}(a)] and $T = 0.8$ [Fig.~\ref{fig:1_h1_quench}(b)]. The system--bath coupling is chosen such that the de-excitation rate for high-energy transitions, $\gamma^{\Downarrow}_{\mathrm{high}}=0.05$, is $r$ times larger than the corresponding low-energy rate $\gamma^{\Downarrow}_{\mathrm{low}}$. We consider two cases: $r=10$ (left panels) and $r=20$ (right panels). Notably, the order parameter starts from a finite initial value, $m_{x,1}(0)=0.71$ in Fig.~\ref{fig:1_h1_quench}(a) and $m_{x,1}(0)=0.12$ in Fig.~\ref{fig:1_h1_quench}(b). Following the quench, $m_{x,1}(t)$  exhibits damped oscillations and ultimately relaxes to zero, indicating relaxation into the disordered phase.  As $r$ increases, relaxation within the low-energy sector is suppressed, effectively restricting the process to a single sector, as described in Eq.~\eqref{assumption_r}. Consequently, increasing the ratio $r$ decreases the damping rate, leading to a slower approach to the asymptotic state.

\begin{figure}[ht!]
    \centering

    \begin{tabular}{@{}cc@{}}
             \includegraphics[width=0.46\linewidth]{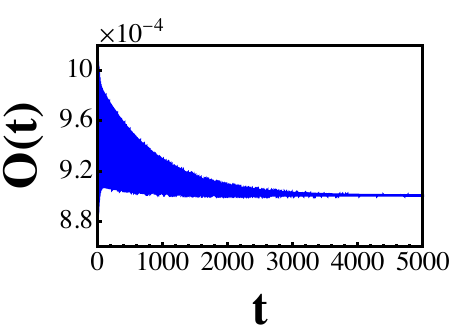} &
			\includegraphics[width=0.46\linewidth]{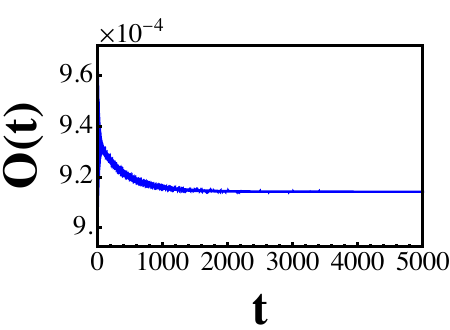}
			\end{tabular}
    \caption{Post quench time dependence of $O(t)$ for fixed $T =0.05$ (left) and $T =0.8$ (right). At $t=0$, a sudden quench of $h_1$ is performed  from $h_{1, i}$ = 0.7 to $h_{1, f} =1.1$. In both the cases $r=20$. }
    \label{fig:1_O_h1_quench}
\end{figure}
\noindent
We also note that, while  the asymptotic, post-quench steady state lies in the disordered phase, with $m_{x,1}=m_{x,2}=0$, the inter-model correlator remains finite, due to the nonzero coupling $\lambda$. This shows how the two models still remain correlated with each other, that is, $\langle \sigma^x_1 \sigma^x_2 \rangle \neq \langle \sigma^x_1 \rangle \langle \sigma^x_2 \rangle$. In Fig.~\ref{fig:1_O_h1_quench}, we plot the time evolution of $O(t)$ following a quench of $h_1$ from $h_{1,i} = 0.7$ to $h_{1,f} = 1.1$, with the temperature fixed at $T = 0.05$ (left) and $T = 0.8$ (right). At $t=0$, $O(0)$   starts from a finite value and, after the quench, it exhibits damped oscillations, before relaxing to a finite steady-state value. Because $\lambda$ is small, $O(t)$ remains weak, but dissipation is not sufficient to fully erase the residual inter-model correlations. At higher temperatures, the oscillations are further suppressed due to enhanced thermal fluctuations. 

To further characterize the relaxation dynamics, we compute the time dependent mixed-state fidelity per site,
\begin{equation}
\mathcal{F}(t) = \left[ \mathrm{Tr} \left( \sqrt{ \sqrt{\hat{\rho}_i(0)}\,\hat{\rho}_i (t)\,\sqrt{\hat{\rho}_i (0)} } \right) \right]^2
\:\:.
\label{eq:fidelity}
\end{equation}
\noindent
The fidelity satisfies the inequality $0 \le \mathcal{F}(t) \le 1$, with $\mathcal{F}(0)=1$. Since the mean-field density matrix factorizes over lattice sites in the thermodynamic limit,  $\mathcal{F}(t)$  captures the local inter-model correlations.
\begin{figure}[ht!]
	\centering
	\begin{tikzpicture}
\node (A) at (0,0) {
\begin{tabular}{@{}cc@{}}
\includegraphics[width=0.46\linewidth]{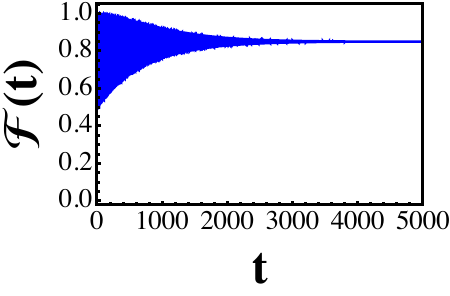} &
\includegraphics[width=0.46\linewidth]{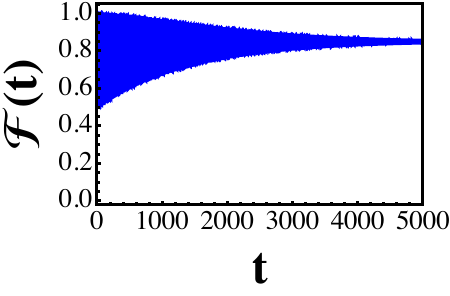}
\end{tabular}
};
\node at (A.north west) [anchor=north west, xshift=-4pt, yshift=-2pt] {\textbf{(a)}};
\end{tikzpicture}

\vspace{0.3cm}

\begin{tikzpicture}
\node (B) at (0,0) {
\begin{tabular}{@{}cc@{}}
\includegraphics[width=0.46\linewidth]{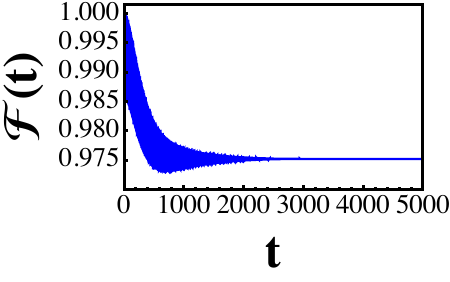} &
\includegraphics[width=0.46\linewidth]{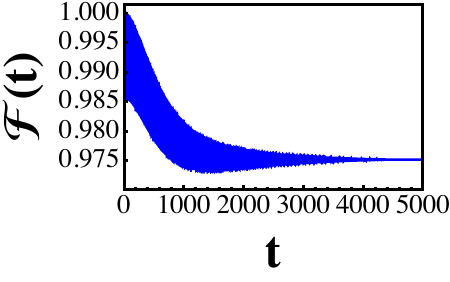}
\end{tabular}
};
\node at (B.north west) [anchor=north west, xshift=-4pt, yshift=-2pt] {\textbf{(b)}};
\end{tikzpicture}
	\caption{Post-quench fidelity per site $\mathcal{F}(t)$ (see text)
	for $r=10$ (left) and $r=20$ (right), for the same quench protocol as in Fig.~\ref{fig:1_h1_quench}.}
	\label{fig:2_h1_quench}
\end{figure}
 
Figure~\ref{fig:2_h1_quench} shows $\mathcal{F} (t)$  as a function of time, computed for the same quench protocol as in Fig.~\ref{fig:1_h1_quench}, with the bath temperature fixed at $T = 0.05$ (\textbf{(a)}) and $T = 0.8$ (\textbf{(b)}). The damping rate depends on the ratio $r$. Specifically, $\mathcal{F}(t)$ damps faster in the left panels ($r=10$) than in the right panels ($r=20$). The relaxation dynamics depends strongly on the temperature. At higher temperatures, the oscillations are suppressed and the fidelity saturates at a higher steady-state value.  Note that, the fidelity per site starts from $\mathcal{F}(0)=1$ and relaxes to a finite value smaller than unity. Given that $\hat{\rho}_i (t)$  deviates significantly from the its initial value $\hat{\rho}_i (0)$, one might expect that $\mathcal{F} (t)$  decays to zero. However, because the total density  matrix factorizes into single-site density matrices, the  total fidelity scales as $F_{\text{total}} \sim (\mathcal{F})^N$. Consequently, in the thermodynamic limit ($N \to \infty$), the total Fidelity indeed vanishes  as long as the Fidelity per site remains strictly less than unity. The finite long-time value of $\mathcal{F}(t)$ indicates that the single-site steady-state density matrix is not orthogonal to the initial state.  In contrast to fidelity-based dynamical phase transitions in closed \cite{Heyl2019, Sehrawat_2021} and open quantum systems \cite{Nava2023_s,Nava2024_s,Wu2024, Nava2025_s}, where nonanalytic cusps appear at critical times, the fidelity here exhibits smooth, damped oscillations and saturates to a finite value at long times. Dissipation therefore smooths out any nonanalytic features in the fidelity. Even when decreasing the dissipation rate $\gamma$, the features of the unitary case do not reemerge. Instead, the system simply requires more time to settle into a steady-state value. The damping rate decreases with increasing $r$, consistent with faster relaxation of high-energy excitations. At low bath temperature, $T = 0.05$, for quenches deep in the disordered phase $h_1> h_{1,c}$, the strong local field suppresses thermal excitations. Therefore, dissipation drives the system close to a unique field polarized product state, yielding an almost pure steady state with $\mathrm{Tr}(\rho_{\mathrm{ss}}^2) \approx 1$. 
   
 It is worth mentioning that when all four energy levels become comparable, as in Eq.~\eqref{comparable_energies}, the results do not reveal any additional qualitatively new features. In this regime, decreasing $h_2$ and increasing $\lambda$ enhances the mean-field feedback, which consequently shifts the equilibrium phase transition line. However, in the presence of dissipation, the relaxation dynamics remain qualitatively similar, with the relaxation time scales primarily governed by  $\gamma$ and by $r$. 
 
So far, we have focused on quenches of  $h_1$, while keeping the bath temperature fixed. We now turn to temperature quenches while keeping $h_1$ fixed. As we show below, in this case the relaxation behavior qualitatively changes and reveals signatures of dynamical phase transitions. The system is again initialized in the ordered phase by choosing appropriate pre-quench parameters, as described in Appendix~\ref{Initial_thermal}. At $t=0$, we perform a temperature quench by suddenly increasing the bath temperature, thereby driving the system into the disordered regime. The post-quench dynamics is governed by the SCMF LME, Eqs.~\eqref{eq:nonlinear_lindblad}, \eqref{eq:Main_MF_H}, and \eqref{eq:m1} with dissipation strength  $\gamma^{\Downarrow}_{\mathrm{high}} \equiv \gamma = r\,\gamma^{\Downarrow}_{\mathrm{low}}$. The initial thermal state serves only as the initial condition, and the system evolves under Lindblad dynamics determined by the post-quench Hamiltonian and bath temperature. As a result, physical observables such as $m_{x,1}(t)$ and $O(t)$ exhibit pronounced nonequilibrium relaxation dynamics. 

\begin{figure}[ht!]
    \centering

    \centering
    \includegraphics[width=0.85\linewidth]{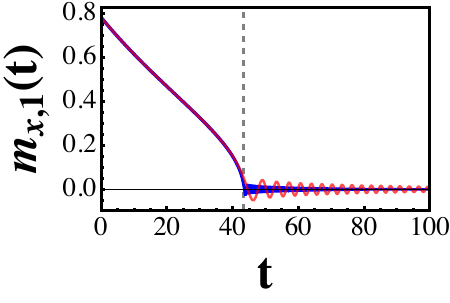}

    \vspace{0.2cm}

    \includegraphics[width=0.85\linewidth]{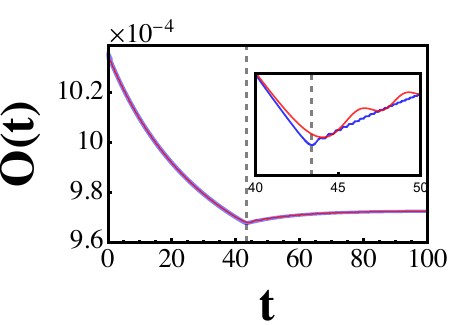}

   \caption{Top: $m_{x,1}(t)$ vs $t\gamma$. Bottom: $O(t)$ vs $t\gamma$. In both panels, we consider a temperature quench $T_i = 0.4 \to T_f = 1.05$ at fixed $h_1 = 0.6$. The blue and red curves correspond to $\gamma = 0.02$ and $\gamma = 0.2$, respectively. Other parameters are $h_2 = 10$, $\lambda = 0.01$, and $r = 20$. The black vertical dashed line indicates the location of the dynamical phase transition. Inset (Bottom panel): $O(t)$ vs $t\gamma$ in the vicinity of the dynamical phase transition.}
    \label{fig:3_T_quench}
\end{figure}
 The top panel of Fig.~\ref{fig:3_T_quench} displays the time evolution of $m_{x,1}(t)$ following a temperature quench, plotted as a function of the rescaled time $t\gamma$. Since, the relaxation time scale depends on the strength of the system-bath coupling, we use this dimensionless time to compare dynamics across different values of $\gamma$. In this plot, we set the transverse field to $h_1 = 0.6$ and initialize the system in the ordered phase at $T_i = 0.4 < T_c(h_1)$. At $t=0$, the temperature is suddenly increased to $T_f = 1.05 > T_c(h_1)$, driving the system into the disordered phase, where $m_{x,1}(t)$ relaxes to zero. Unlike transverse-field quenches, this temperature quench does not induce pronounced initial oscillations. Instead, the dynamics exhibit a dynamical phase transition (DPT) at a critical rescaled time $t_* \gamma$. For low dissipation strengths, this transition is characterized by a sharp, non-analytic change in the relaxation behavior. Notably, the critical rescaled time $t_* \gamma$ is independent of $\gamma$ for all cases exhibiting the DPT. However, as the dissipation strength increases (e.g., $\gamma = 0.2$, red line), this non-analytic behavior is smeared out, and the dynamics exhibit a smooth crossover rather than a sharp transition at the critical rescaled time \cite{Parez:2025_DPT}.

We also investigate the post-quench dynamics of $O(t)$. The bottom panel of Fig.~\ref{fig:3_T_quench} shows $O(t)$ as a function of the rescaled time $t\gamma$ following a temperature quench. We consider the same quench protocol as in the top panel. For  $\gamma = 0.02$ (blue curve), the inter-model correlator exhibits a pronounced kink at the dynamical phase transition time $t_* \gamma$. This nonanalytic feature becomes progressively smeared out for larger $\gamma$. Nevertheless, as in the case of the $h_1$ quench, the steady state reached after the temperature quench retains  residual inter-model correlations.  
     
\begin{figure}[ht!]
	\centering
	\includegraphics[width=0.85\linewidth]{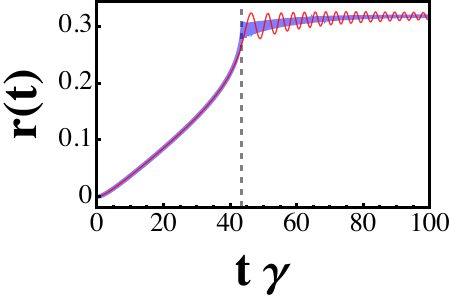}
	\caption{Rate function $r(t)$ as a function of the rescaled time $t\gamma$ for a temperature quench $T_i = 0.4 \to T_f = 1.05$ at fixed $h_1 = 0.6$. The blue and red curves correspond to $\gamma = 0.02$ and $\gamma = 0.2$, respectively. Other parameters are $h_2 = 10$, $\lambda = 0.01$, and $r = 20$. The black vertical dashed line marks the location of the dynamical phase transition.}
	\label{fig:4_T_quench}
\end{figure}
 For a fixed $h_1$, we further characterize the relaxation dynamics following a temperature quench by computing the time-dependent mixed-state fidelity per site using Eq.~\eqref{eq:fidelity}. To probe dynamical phase transitions, we introduce the rate function
\begin{equation}
r(t) = -\log\big(\mathcal{F}(t)\big).
\end{equation}
Figure~\ref{fig:4_T_quench} displays the corresponding dynamics of $r(t)$ for the same quench protocol as in Fig.~\ref{fig:3_T_quench}. For weak dissipation ($\gamma = 0.02$, blue curve), the rate function exhibits a distinct nonanalyticity at a critical time $t_*$ (scaled by $\gamma$), signaling a dynamical phase transition. In contrast, for stronger dissipation, the nonanalytic feature is smoothed out, and the dynamics display a continuous crossover. Unlike the transverse-field quench, here, nonanalytic behavior in the time evolution of observables persists in the presence of dissipation. However, increasing the dissipation strength progressively smears out the dynamical phase transition~\cite{Parez:2025_DPT}. 

 In conclusion, our results demonstrate that, within the time-dependent self-consistent LME framework, when the jump operators are defined in the instantaneous eigenbasis of the mean-field Hamiltonian, the system is driven toward an equilibrium-like state determined by the post-quench parameters. Specifically, it relaxes to an asymptotic steady state that coincides with the thermal Gibbs state associated with the mean-field Hamiltonian. The relaxation dynamics, however, depends strongly on the quench protocol. In the presence of dissipation, a transverse-field quench leads to conventional relaxation without signatures of dynamical phase transitions. In contrast, a temperature quench induces a dynamical phase transition at a critical time scale, manifested as a  nonanalytic change in the time evolution of observables.

 Having established the relaxation dynamics toward the steady state, we now turn to the characterization of the dissipative dynamics described by jump operators defined in the instantaneous eigenbasis of the mean-field Hamiltonian as Eq.~\eqref{jump_eigen}, with transition rates satisfying detailed balance as Eq.~\eqref{eq:detailed_balance}. 

\section{Steady state phase diagram and dissipative phase transition with self-consistently determined dissipators}
\label{sec:DPT_A}
 
In equilibrium statistical mechanics, a second-order phase transition is marked by a continuous order parameter and by a nonanalytic behavior of the response functions. An analogous phenomenology arises in the steady state of open quantum systems, to which the system is driven by the pertinent dissipation mechanism. Typically, a continuous (second-order) DPT is evidenced by  nonanalyticies in the physical properties of the steady state in terms of the control parameters, combined with a corresponding continuous vanishing of the order parameter at criticality \cite{Colli2006}. In our data, the steady-state magnetization decreases smoothly and approaches zero at $h_{1,c}$, and the forward and backward sweeps coincide within numerical accuracy, signaling the absence of a hysteresis loop. This rules out a first-order DPT in the explored parameter regime, where one would expect coexistence of two stable steady states and a hysteretic response under forward and backward sweeps. The observed behavior is therefore consistent with a continuous DPT: the ordered phase ($h_1<h_{1,c}$) is characterized by a nonzero magnetization, whereas the disordered phase ($h_1>h_{1,c}$) displays vanishing magnetization. Accordingly, we characterize the  DPT  by tracking the density matrix operator of the steady state, $\hat{\rho}_{\mathrm{ss}}(h_1)$, as a function of the control parameter $h_1$, while keeping $h_2 = 10$ and $\lambda = 0.01$ fixed, such that the assumption in Eq.~\eqref{assumption_r} is satisfied and, in particular, such that the equilibrium phase diagram remains approximately the same as that shown in Fig.~\ref{fig:1Ising_phase_diagram}. We consider $r=10$ and $\gamma=0.05$,  with all parameters given in units of $J=1$. For each val
ue of $h_1$, the steady state is obtained numerically by evolving the master equation from an initial condition in the ordered phase until convergence. To test for hysteresis and multi-stability we perform both a forward sweep ($h_1$ increasing) and a backward sweep ($h_1$ decreasing), using the converged steady state at one parameter value as the initial condition for the next, a procedure commonly referred to as parameter continuation \cite{Book_continuation_methods, Book_bifurcation}.  
  
To diagnose the transition we compute steady-state observables directly from the asymptotic steady state  density matrix $\hat{\rho}_{\mathrm{ss}}$. Specifically, we use the magnetization in the steady state, $m_{1,\rm{ss}} = \rm{Tr}\big(\hat{\rho}_{\rm{ss}} \sigma^x_{1}\big)$, as the order parameter to map out the DPT, and also monitor the inter-model correlator  $O_{\mathrm{ss}} = \rm{Tr}\big(\hat{\rho}_{\rm{ss}} \sigma^x_{1} \sigma^x_{2}\big)$, which provides complementary information on steady-state correlations between the two QIMs across the transition. 

\begin{figure}[ht!]
    \centering
    \begin{tabular}{@{}cc@{}}
			\includegraphics[width=0.46\linewidth]{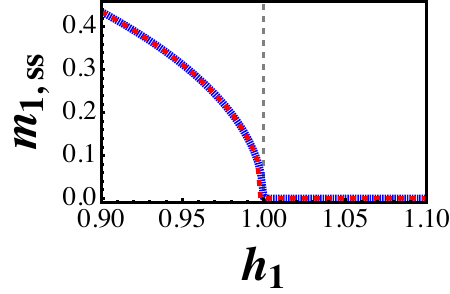} &
			\includegraphics[width=0.46\linewidth]{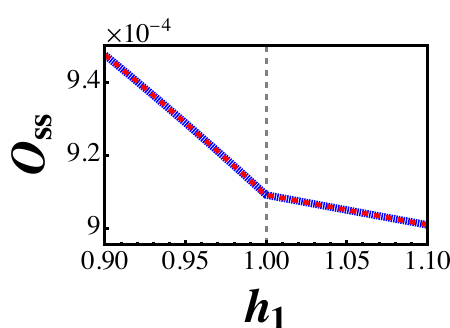}
			\end{tabular}
    \caption{Steady state magnetization $m_{1, \rm{ss}}$ (left) and inter-model correlator $O_{\rm{ss}}$ (right) for forward (dashed blue) and backward sweep (dashed red) of $h_1$. The black vertical line indicates the transition line at $h_{1,c} =1$. We keep bath temperature fixed at $T=0.05$.}
    \label{fig:m1_O_h1_sweep}
\end{figure}
In Figure~\ref{fig:m1_O_h1_sweep}, we plot $m_{1, \rm{ss}}$ and $O_{\rm{ss}}$ as functions of $h_1$ for forward and backward sweep of $h_1$ in dashed blue and dashed red, respectively. The magnetization $m_{1, \rm{ss}}$ remains non zero for $h_1< h_{1,c}$ and decreases smoothly to zero at $h_{1,c} \approx 1$, with no discernible hysteresis. The correlator $O_{\rm{ss}}$ remains continuous across the transition, while exhibiting a change in slope near $h_{1,c}$ and shows no hysteretic behavior as well. These observations are consistent with a continuous DPT. To further monitor the DPT, we also look at the expected enhancement of the quantum correlations for $h_1 \sim h_{1,c}$. Indeed, in driven-dissipative systems, quantum correlations in mixed states are often enhanced near critical points, as evidenced by a peak in the logarithmic entanglement negativity \cite{Casteels:2016ahe}. To spell out this point,  beyond local order parameter,  we also look at quantum correlations in the steady state, by computing the concurrence $C(\rho_{\rm{ss}})$, a standard entanglement monotone for two-qubit mixed states. Concurrence measures how far a two-qubit state deviates from separability, with $C=0$ for separable states and $C=1$ for maximally entangled Bell states. For a two-qubit density matrix $\rho$, the concurrence is defined as
\begin{equation}
C(\rho)=\max\!\left\{0,\, \lambda_1-\lambda_2-\lambda_3-\lambda_4\right\}
\:\:, 
\label{eq:concurrence}
\end{equation}
\noindent
where $\lambda_i$ are the square roots of the eigenvalues of the matrix $\hat{\rho} \tilde{\hat{\rho}}$ sorted in descending order, with
\begin{equation}
\tilde{\hat{\rho}}= (\sigma_y\otimes\sigma_y)\,\tilde{\rho}^*\,(\sigma_y\otimes\sigma_y)\:\:, 
\label{eq:spinflip}
\end{equation}
\noindent
with $\tilde{\rho}^*$ being the complex conjugate of $\hat{\rho}$ and $\sigma_y$ being  the Pauli-$y$ matrix. 
 
\begin{figure}[ht!]
    \centering
    \begin{tabular}{@{}cc@{}}
			\includegraphics[width=0.46\linewidth]{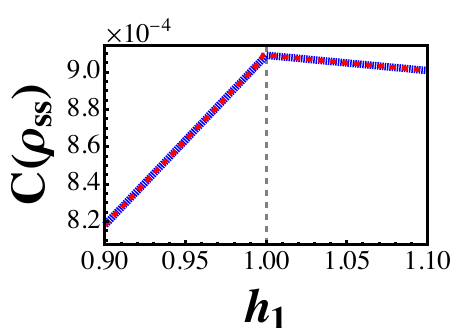} &
			\includegraphics[width=0.46\linewidth]{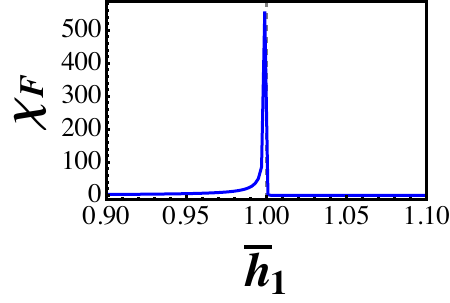}
			\end{tabular}
    \caption{Left: NESS concurrence $C(\rho_{\rm{ss}})$ for forward (dashed blue) and backward sweep (dashed red) of $h_1$. The black vertical line indicates the transition line at $h_{1,c} =1$. Right : Fidelity susceptibility $\chi_F$ evaluated at the mid points of consecutive $h_1$ values as $\bar{h}_1 = \frac{h_{1, i}+ h_{1, i+1}}{2}$. We keep bath temperature fixed at $T=0.05$.}
    \label{fig:C_chiF}
\end{figure}
In the left panel of Fig.~\ref{fig:C_chiF}, we plot $C(\rho_{\rm{ss}})$ for forward and backward sweeps of $h_1$, shown by dashed blue and dashed red lines, respectively. Here as well, no hysteresis is observed. The concurrence exhibits a peak-like enhancement near the continuous transition points $h_{1,c}$, reflecting the build-up of non-classical correlations at criticality.

To probe the sensitivity of the steady state to changes in $h_1$, we compute the fidelity susceptibility $\chi_F$, defined via the Uhlmann fidelity between neighboring steady states. The Uhlmann fidelity is $F(\rho,\sigma)= \left[\mathrm{Tr}\sqrt{\sqrt{\rho}\,\sigma\,\sqrt{\rho}}\right]^2$,  
and the fidelity susceptibility is obtained as the second-order expansion \cite{PhysRevA.30.1610, PhysRevA.71.032313, PhysRevE.90.052103}
\begin{equation}
\chi_F(h_1) =\lim_{\delta h\to 0}\, \frac{2\big[1-F\big(\hat{\rho}_{\mathrm{ss}}(h_1),\hat{\rho}_{\mathrm{ss}}(h_1+\delta h)\big)\big]}{\delta h^2}
\:\: . \label{eq:fidsus}
\end{equation}
\noindent
(Here, the Uhlmann fidelity is evaluated between steady-state density matrices at neighboring values of $h_1$, and is therefore time independent.) From a quantum information perspective, $\chi_F$ is directly proportional to the quantum Fisher information, and thus  quantifies the enhanced parameter sensitivity near the transition. Indeed, as shown in the right panel of Figure~\ref{fig:C_chiF},  $\chi_F$ exhibits a pronounced nonanalytic spike at the critical point, signaling a continuous dissipative phase transition through enhanced steady-state sensitivity \cite{PhysRevA.105.052226}. A pronounced peak in $\chi_F$ indicates that $\rho_{\mathrm{ss}}$ changes rapidly with the control parameter and is a widely used signature of continuous criticality. Notably, although two-spin correlations and weak entanglement persist for $h_1>h_{1,c}$ due to the interaction term, the steady state varies smoothly and approximately linearly with the control parameter in this regime, resulting in a vanishing  $\chi_F$.

To further deepen our analysis, we now perform a linear stability analysis of the asymptotic steady state. Indeed, while the behavior of local observables and fidelity provides a macroscopic signature of the transition, a deeper understanding of the steady-state stability is obtained through linear stability analysis, which eventually allows for characterizing the critical slowing down near the transition and for identifying  the softening of the relaxation gap that signals the onset of the phase transition \cite{Debecker_2025, Debecker_2024, Kothe:2024}. In a compact form, we rewrite the LME for $\hat{|rho} (t)$ in a Markovian,  open quantum system in the form  
\begin{equation}
\frac{d \hat{\rho}(t)}{d t} = {\bf L}_t [\hat{\rho}(t)] 
\:\: , 
\label{lme.1}
\end{equation}
where ${\bf L}_t$ is a superoperator acting on the density operator, i.e., a (generally nonlinear) functional of $\hat{\rho}$. In many cases of physical interest, ${\bf L}_t$ is a linear functional, and Eq.~(\ref{lme.1}) reduces to 
\begin{equation}
\frac{d \hat{\rho}(t)}{d t} = {\cal L} \hat{\rho}(t) 
\;\; ,
\label{lme.2}
\end{equation}
where ${\cal L}$ is the Liouvillian superoperator. In general, the relaxation toward the steady state is determined by the spectral properties of $\mathcal{L}$. Denoting by $\{\lambda_i(\mathcal{L})\}$ its eigenvalues, the Liouvillian gap
\begin{equation}
\Delta_L = -\max_{\lambda_i \neq 0} \Re\,\lambda_i(\mathcal{L})
\;\; ,
\label{ligap}
\end{equation}
sets the longest relaxation time, $\tau \sim 1/\Delta_L$. In linear dissipative systems, the closing of $\Delta_L$ at continuous dissipative phase transitions signals critical slowing down and the emergence of a soft mode 
\cite{PhysRevA.86.012116, PhysRevA.98.042118}.

In contrast, the dynamics within our time-dependent mean-field approximation is governed by the \emph{nonlinear} LME in Eq.~\eqref{eq:nonlinear_lindblad}, where both the effective Hamiltonian and the dissipative processes depend self-consistently on the instantaneous density matrix. As a result, no global linear Liouvillian superoperator exists, and the Liouvillian gap in the strict spectral sense is not well defined. Therefore, to characterize the relaxation properties in this case, we resort to a \emph{linear stability analysis} (LSA) around the asymptotic steady state $\hat{\rho}_{\mathrm{ss}}$. By definition, $\hat{\rho}_{\mathrm{ss}}$ satisfies ${\bf L}_t[\hat{\rho}_{\mathrm{ss}}] = 0$. We then consider a small time-dependent perturbation $\delta \hat{\rho}(t)$, such that $\hat{\rho}(t) = \hat{\rho}_{\mathrm{ss}} + \delta \hat{\rho}(t)$. For convenience, we implicitly vectorize the density matrix and treat the Lindblad equation as a linear evolution in operator space. Expanding Eq.~(\ref{lme.1}) to first order in $\delta \hat{\rho}(t)$, we obtain
\begin{equation}
\frac{d\,\delta\hat{\rho}(t)}{dt} = J[\hat{\rho}_{\mathrm{ss}}] \, 
\delta\hat{\rho}(t)
\;\; ,
\end{equation}
where the Jacobian
\begin{equation}
J = \left.\frac{\partial \mathbf{L}_t[\hat{\rho}]}{\partial \hat{\rho}}\right|_{\hat{\rho}=\hat{\rho}_{\mathrm{ss}}}
\;\; ,
\label{jacobian}
\end{equation}
governs the local relaxation dynamics.

In general, $J$ has a complex spectrum. Let $\{\mu_i\} = \{\alpha_i + i\omega_i\}$ be its eigenvalues, which determine the decay rates and oscillatory components of perturbations, with $\alpha_i < 0$ ensuring dynamical stability. We define the linear stability gap as
\begin{equation}
\Delta_{\mathrm{stab}} = -\max_i \Re(\mu_i)
\;\; ,
\label{eq:stab_gap}
\end{equation}
which corresponds to the slowest local decay rate and sets the longest local relaxation time $\tau_{\mathrm{stab}} \sim 1/\Delta_{\mathrm{stab}}$. The imaginary part $\omega_{\mathrm{slow}}$ of the corresponding eigenvalue characterizes possible damped oscillatory relaxation. While a vanishing real part with finite $\omega_{\mathrm{slow}}$ would indicate a Hopf bifurcation and the onset of limit-cycle dynamics, all eigenvalues in the parameter regime considered here satisfy $\mathrm{Re}(\mu_i) < 0$, thereby excluding such instabilities. At this point, it is important to emphasize that DPTs are global phenomena, characterized by nonanalytic changes of the steady-state manifold. LSA, by contrast, only probes local relaxation properties, in the vicinity of a given steady state. Consequently, while $\Delta_{\mathrm{stab}}$ provides valuable information about local relaxation dynamics around a given steady state, it does not by itself determine which steady state is dynamically selected or capture nonanalytic changes of the steady-state manifold. Thus, LSA does not constitute a universal diagnostic of continuous DPTs.

\begin{figure}
    \centering
    \begin{tabular}{@{}cc@{}}
        \includegraphics[width=0.46\linewidth]{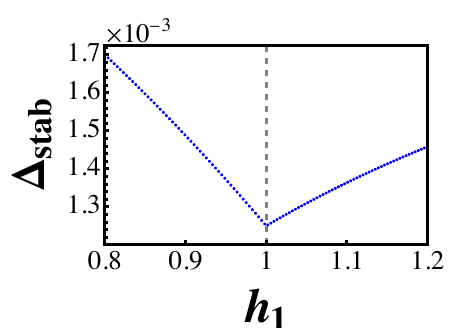} &
        \includegraphics[width=0.46\linewidth]{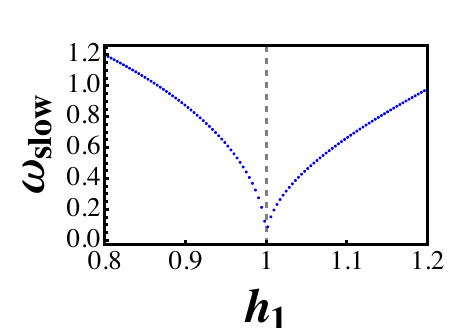}
    \end{tabular}
    \caption{Left: Linear stability gap $\Delta_{\mathrm{stab}}$ as a function of $h_1$. 
    Right: Imaginary part $\omega_{\rm slow}$ of the slowest decaying mode. In both the plots, we keep $h_2=10$ and $\lambda=0.01$.}
    \label{fig:LSA_h1}
\end{figure}
To fully characterize the transition, we therefore complement the linear stability analysis with a direct study of the relaxation dynamics. For each value of $h_1$, the steady state is obtained numerically by evolving the master equation from an initial condition within the ordered phase until convergence. We perform a forward sweep over $h_1$ using the parameter continuation procedure described previously. Upon reaching the steady state for a given $h_1$, we introduce a small perturbation and compute the characteristic time scales required for the system to relax back to the steady state corresponding to those parameter values. As shown in Fig.~\ref{fig:LSA_h1}, the stability gap remains finite and positive across the entire parameter range of $h_1$, indicating the absence of true critical slowing down. This implies that the numerical time evolution converges to a stable steady state for all values of $h_1$.  Nevertheless, $\Delta_{\mathrm{stab}}$ exhibits a pronounced cusp-like minimum near the critical point $h_{1,c}$, reflecting a significant slowing down of local relaxation dynamics in the vicinity of criticality. At the same time, the imaginary part $\omega_{\rm slow}$ of the slowest mode displays a corresponding minimum, indicating a softening of the dominant oscillatory mode near the transition without the onset of limit-cycle dynamics. Deep in the disordered phase, the gap increases again, consistent with the emergence of a strongly attractive, field-polarized steady state.

We therefore conclude that while linear stability analysis does not replace global steady-state diagnostics such as order parameters or fidelity susceptibility, it provides a complementary dynamical perspective on dissipative criticality by quantifying how perturbations relax near the steady state.

In summary, the continuous vanishing of the order parameter $m_{1,ss}$ at the critical point $h_{1,c}$, coupled with the absence of hysteresis during forward and backward parameter sweeps, identifies this transition as a continuous (second-order) dissipative phase transition (DPT). To characterize this transition, we employ the concurrence and fidelity susceptibility as complementary diagnostics. The concurrence serves to quantify steady-state entanglement, which is typically enhanced in the vicinity of the critical point. Simultaneously, the fidelity susceptibility exhibits a pronounced peak at $h_{1,c}$, capturing the heightened sensitivity of the asymptotic steady state to infinitesimal parameter variations.  Crucially, our choice of dissipative channels, which induce transitions satisfying detailed balance between the instantaneous eigenstates of the mean-field Hamiltonian, ensures that the asymptotic steady state remains nearly pure at low temperatures. Under these conditions, the steady state closely follows the equilibrium mean-field ground state. Consequently, the dissipative phase transition is effectively pinned to the equilibrium critical point and is characterized by nonanalytic changes in steady-state observables, notably occurring without the closing of the Liouvillian gap. Incidentally, one might wonder whether the equilibrium-like behavior observed in this case is simply a consequence of the weak system–bath coupling. However, this is not the primary origin. Instead, it arises from the structure of the dissipative channels, which enforce detailed balance in the instantaneous eigenbasis of the mean-field Hamiltonian. This drives the system toward a steady state closely related to the equilibrium ground state at low temperature. We have verified that, increasing the system–bath coupling, while remaining within the regime of validity of the Lindblad master equation, primarily affects the relaxation time scales, leading to faster convergence toward the steady state, without altering the structure of the steady state itself (see also \cite{Nava2025_s, Giuliano2026} for a detailed discussion about this point). 
 
In the next section, we study the DPT  using local spin raising and lowering operators as jump operators. Since these jump operators are not defined in the instantaneous eigenbasis of the Hamiltonian, the dissipation is no longer self-consistently aligned with the mean-field Hamiltonian, although the Hamiltonian itself remains self-consistently determined. In this case, the DPT  arises from the competition between the unitary dynamics and the dissipation, and the critical point of the dissipative phase transition is not necessarily the same as at the equilibrium critical point.

\section{Dissipative phase transition with local pump-loss dissipation}
\label{DPT_B}

In this section, we modify the set of dissipators entering the Lindblad master equation (LME) for the density matrix of the system. In particular, while the unitary evolution of the single-site density matrix, $\hat{\rho}_i \equiv \hat{\rho}$, is still determined by the mean-field Hamiltonian $\hat{H}_{\mathrm{MF},i} \equiv \hat{H}_{\mathrm{MF}}$ in Eq.~\eqref{eq:Main_MF_H}, we alter the dissipative term such that the LME  takes the form
\begin{align}
\frac{d \hat{\rho}(t)}{dt} &= -i[\hat H_{\mathrm{MF}}(\hat{\rho}(t)), \hat{\rho}(t)] 
+ \gamma_\downarrow \sum_{n=1}^{2} \mathcal{D}[\sigma_n^-]\hat{\rho}(t) \nonumber \\
&\quad + \gamma_\uparrow \sum_{n=1}^{2} \mathcal{D}[\sigma_n^+]\hat{\rho}(t) \:\:,
\label{LME_local}
\end{align}
\noindent
with 
\begin{equation}
\mathcal{D}[L]\hat{\rho} = L\hat{\rho} L^\dagger -\frac{1}{2}\{L^\dagger L,\hat{\rho}\}
\:\:.
\label{dissipators}
\end{equation}
\noindent
The dissipators constructed from the local spin raising and lowering operators are defined as  
\begin{equation}
\sigma_n^{\pm} = \frac{1}{2}\left(\sigma_{n}^{x} \pm i \sigma_{n}^{y}\right)
\:\:.
\label{rais.lower}
\end{equation}
\noindent
 The rate $\gamma_{\uparrow}$ denotes the incoherent pumping rate associated with the excitation process generated by $\sigma^+$, while $\gamma_{\downarrow}$ denotes the loss rate associated with relaxation generated by $\sigma^-$.  Note that in the previous case (Sec.~\ref{sec:DPT_A}), the jump operators were defined in the instantaneous eigenbasis of the mean-field Hamiltonian and, therefore, they  depended self-consistently on the state of the system. This leads to a nonlinear dissipative dynamics that drives the system toward a steady state closely related to the mean-field ground state. In the present case, we instead consider local spin raising and lowering operators as jump operators, which generate a genuine nonequilibrium steady state in which the phase transition arises from the competition between coherent dynamics and dissipation. 

To analyze the DPTs in terms of physically relevant observables, we derive a closed set of equations of motion for their expectation values. Exploiting translational invariance of the self-consistent solution, we drop the site index $_i$ and define the expectation value of a local observable $\hat{O}$ as
\begin{equation}
  O(t)  = \mathrm{Tr}\!\left(\hat{O} \hat{\rho} (t)\right)
\:\:.
\label{exva}
\end{equation}
\noindent
Using the LME, we recover the equation of motion for   $O (t)$ in the form 
\begin{eqnarray}
\frac{d O (t)}{dt}  
&=& -i\langle [\hat{O}, H_{\mathrm{MF}}]\rangle \nonumber \\
&+& \gamma_\downarrow \sum_{n=1}^{2} 
\left\langle \sigma_n^+ \hat{O} \sigma_n^- 
- \frac{1}{2}\{\sigma_n^+\sigma_n^-, \hat{O}\} \right\rangle \nonumber \\
&+& \gamma_\uparrow \sum_{n=1}^{2} 
\left\langle \sigma_n^- \hat{O} \sigma_n^+ 
- \frac{1}{2}\{\sigma_n^-\sigma_n^+, \hat{O}\} \right\rangle
\:\:.
\label{exva.2}
\end{eqnarray}
\noindent
As in the case of the self-consistently defined dissipators, we define the local magnetizations as 
\begin{equation}
m_{\alpha,n} = \langle \sigma_n^\alpha \rangle 
\;\; , 
\qquad \alpha = x,y,z 
\;\;,
\end{equation}
\noindent 
and the inter-model correlators as 
\begin{equation}
O_{\alpha\beta} = \langle \sigma_1^\alpha \sigma_2^\beta \rangle 
\:\:.
\end{equation}
\noindent
The dissipative dynamics induces an explicit time dependence in all these observables. The resulting set of equations of motion for the local magnetizations and the inter-model correlators can be directly derived from the LME in Eq.~\eqref{LME_local} and read as

\begin{subequations}
   \begin{align}
    \frac{d m_{x,} (t)}{dt}  &= 2h_n m_{y,n } (t)  -\frac{\gamma_\uparrow+\gamma_\downarrow}{2} m_{x,n} (t) \quad  n= 1,2,\\
        \frac{ d m_{y,1} (t)}{dt}  &= 2\left(m_{x,1} (t) m_{z,1} (t) -h_1 m_{x,1}  (t)+\lambda O_{zx} (t) \right) \nonumber \\
        &-\frac{\gamma_\uparrow+\gamma_\downarrow}{2} m_{y,1} (t), \\
        \frac{ d m_{y,2} (t)}{dt} &= 2\left(m_{x,2} (t) m_{z,2} (t)-h_2 m_{x,2} (t)+\lambda O_{xz} (t) \right) \nonumber \\&- \frac{\gamma_\uparrow+\gamma_\downarrow}{2} m_{y,2} (t), \\
        \frac{d m_{z,1} (t)}{d t}  &= -2\left(m_{x,1}(t) m_{y,1} (t)+\lambda O_{yx}(t) \right) \nonumber \\&+ (\gamma_\uparrow-\gamma_\downarrow) - (\gamma_\uparrow+\gamma_\downarrow)m_{z,1}(t), \\
        \frac{d m_{z,2} (t)}{d t}  &= -2\left(m_{x,2} (t) m_{y,2} (t) +\lambda O_{xy}(t)\right) \nonumber \\&+ (\gamma_\uparrow-\gamma_\downarrow) - (\gamma_\uparrow+\gamma_\downarrow)m_{z,2}(t), \\
    \frac{d  O_{xx} (t)}{dt}  &= 2(h_1 O_{yx}(t)+h_2 O_{xy}(t)) \nonumber \\&-(\gamma_\uparrow+\gamma_\downarrow)O_{xx}(t), \\
\frac{d  O_{yy} (t)}{dt} &= 2(m_{x,1}(t) O_{zy}(t)+m_{x,2}(t) O_{yz}(t) \nonumber \\&-h_1 O_{xy}(t) -h_2 O_{yx}(t)) \nonumber \\&-(\gamma_\uparrow+\gamma_\downarrow)O_{yy}(t), \\
\frac{d O_{zz} (t)}{dt}  &= -2(m_{x,1}(t) O_{yz}(t)+m_{x,2}(t)O_{zy}(t)) \nonumber \\&+(\gamma_\uparrow-\gamma_\downarrow)(m_{z,1}(t) +m_{z,2}(t)) \nonumber \\&-2(\gamma_\uparrow+\gamma_\downarrow)O_{zz}(t), \\
\frac{ d O_{xy}(t)}{dt}  &= 2\big(h_1 O_{yy} (t)+ m_{x,2}(t) O_{xz} (t) - h_2 O_{xx} (t) \nonumber \\ &+ \lambda m_{z,2}(t)\big) - (\gamma_{\uparrow} + \gamma_{\downarrow}) O_{xy}(t), \\
\frac{ d O_{xz} (t)}{dt}  &= 2 \big( h_1 O_{yz}(t) - m_{x,2} (t) O_{xy} (t)- \lambda m_{y,2}(t) \big) \nonumber \\&+ (\gamma_{\uparrow} - \gamma_{\downarrow}) m_{x,1} (t) - \frac{3}{2}(\gamma_{\uparrow} + \gamma_{\downarrow}) O_{xz} (t) , \\
\frac{ d O_{yx} (t)}{dt} &= 2m_{x,1} (t)O_{zx}(t) - 2h_1 O_{xx}(t) + 2h_2 O_{yy} (t) \nonumber \\&+ 2\lambda m_{z,1}(t) - (\gamma_{\uparrow} + \gamma_{\downarrow}) O_{yx}(t),\\
\frac{ d O_{yz} (t)}{dt}  &= 2\big( m_{x,1} (t)O_{zz}(t) - h_1 O_{xz} (t)\nonumber \\&- m_{x,2}(t) O_{yy} (t)\big)+ (\gamma_{\uparrow} - \gamma_{\downarrow}) m_{y,1}(t) \nonumber \\&- \frac{3}{2}(\gamma_{\uparrow} + \gamma_{\downarrow}) O_{yz}(t), \\
\frac{ d O_{zx}(t)}{d t}  &= -2\big(m_{x,1}(t) O_{yx} (t)- h_2 O_{zy}(t) + \lambda m_{y,1}(t) \big)\nonumber \\& + (\gamma_{\uparrow} - \gamma_{\downarrow}) m_{x,2} (t) - \frac{3}{2}(\gamma_{\uparrow} + \gamma_{\downarrow}) O_{zx} (t),\\
\frac{d O_{zy} (t)}{dt}  &= -2\big(m_{x,1}(t) O_{yy}(t) - m_{x,2}(t) O_{zz} (t) \nonumber \\&+ h_2 O_{zx}(t) \big) + (\gamma_{\uparrow} - \gamma_{\downarrow}) m_{y,2} (t) \nonumber \\&- \frac{3}{2}(\gamma_{\uparrow} + \gamma_{\downarrow}) O_{zy}(t).
\end{align}
\label{eom_all}
\end{subequations}
\noindent
Clearly, all the steady-state solutions are obtained by setting to zero the right-hand side of each one of Eqs.~\eqref{eom_all}, which leads to a system of coupled nonlinear algebraic equations whose solutions determine the nonequilibrium phases of the system \cite{Kothe:2025, Lee_2014,Lee_2023}.

We begin by using Eqs.~\eqref{eom_all} to determine the steady-state solutions in the symmetric (disordered) phase. To this end, we note that the system possesses a global $\mathbb{Z}_2$ symmetry generated by the parity operator
\begin{equation}
\Pi = \sigma^z_1 \sigma^z_2
\:\: .
\end{equation}
\noindent
Under this transformation, the spin operators transform as
\begin{equation}
\Pi \sigma^x_n \Pi = -\sigma^x_n, \qquad 
\Pi \sigma^y_n \Pi = -\sigma^y_n, \qquad 
\Pi \sigma^z_n \Pi = \sigma^z_n
\:\:.
\end{equation}
\noindent
The mean-field Hamiltonian is invariant under the global $\mathbb{Z}_2$ transformation $\Pi = \sigma_1^z \sigma_2^z$ only in a self-consistent sense, since the order parameter  $m_{x,n} = \langle \sigma_n^x \rangle$ changes sign under the transformation. Therefore, the symmetric phase corresponds to solutions with $m_{x,n}=0$, while solutions with $m_{x,n} \neq 0$ correspond to symmetry-broken phases.
The dissipative part of the dynamics remains invariant under this transformation because
\begin{equation}
\Pi \sigma_n^{\pm} \Pi = - \sigma_n^{\pm} \;\;,
\end{equation}
\noindent
and the Lindblad dissipator satisfies $\mathcal{D}[-L] = \mathcal{D}[L]$. Therefore, the full dynamics is invariant under the $\mathbb{Z}_2$ symmetry.

This symmetry allows us to classify observables into parity-even and parity-odd ones. 
In the symmetric phase, all the parity-odd observables must vanish. This implies
\begin{equation}
m_{x,n} = 0, \qquad m_{y,n} = 0, \qquad n = 1,2
\;\;,
\label{obs0}
\end{equation}
\noindent
and the correlators containing an odd number of $x$ or $y$ operators also vanish,
that is

\begin{equation}
O_{xz} = O_{zx} = O_{yz} = O_{zy} = 0
\;\;.
\label{obs1}
\end{equation}
\noindent
In contrast, the observables in the even sector can remain finite. Therefore, we impose the 
symmetric-phase ansatz and solve the reduced set of algebraic equations in the even sector to determine the symmetric steady-state solutions. The resulting reduced set of algebraic equations is  given by
\begin{subequations}
    \begin{align}
   m_{z,1} (\gamma_{\downarrow} + \gamma_{\uparrow})+ \gamma_{\downarrow} -\gamma_{\uparrow}+ 2\lambda  O_{yx}&=0 \\ 
   m_{z,2} (\gamma_{\downarrow}+\gamma_{\uparrow})+\gamma_{\downarrow} - \gamma_{\uparrow}+2 \lambda O_{xy}&=0\\ 
   2 (h_1 O_{yx} + h_2 O_{xy})- O_{xx}(\gamma_{\downarrow}+\gamma_{\uparrow})&=0 \\ 
   2 (h_1 O_{xy} +h_2 O_{yx} )+ O_{yy}( \gamma_{\downarrow}+ \gamma_{\uparrow})&=0 \\
   (\gamma_{\uparrow}- \gamma_{\downarrow}) (m_{z,1}+ m_{z,2})-2 O_{zz} (\gamma_{\downarrow}+ \gamma_{\uparrow})&=0 \\ 
   O_{xy} (\gamma_{\downarrow}+ \gamma_{\uparrow})-2 h_1 O_{yy}+2 h_2 O_{xx} - 2 \lambda m_{z,2} &=0 \\
   O_{yx} (\gamma_{\downarrow}+ \gamma_{\uparrow})+2 h_1 O_{xx} -2 h_2 O_{yy}-2 \lambda  m_{z,1} &=0 \:\: .
\end{align}
\label{eom_reduced}
\end{subequations}
\noindent
Accordingly, the steady-state solution in the symmetric phase, which we denote by the vector $\vec{\nu}_S$ in the following, is given by
\begin{align}
m_{x,n} &= m_{y,n} = 0, \qquad \text{for} \; n=1,2, \nonumber \\
O_{xz} &= O_{zx} = O_{yz} = O_{zy} = 0, \nonumber \\
m_{z, n=1,2} &= -\frac{(\gamma_\downarrow - \gamma_{\uparrow}) \left[(\gamma_\downarrow + \gamma_{\uparrow})^2 + 4 (h_1 + h_2)^2\right]} {(\gamma_\downarrow + \gamma_{\uparrow}) \left[(\gamma_\downarrow + \gamma_{\uparrow})^2 + 4 \left((h_1 + h_2)^2 + \lambda^2\right)\right]}, \nonumber \\
O_{xx} &= -\frac{4 \lambda (\gamma_{\downarrow} - \gamma_{\uparrow})(h_1 + h_2)} {(\gamma_{\downarrow} + \gamma_{\uparrow}) \left[(\gamma_{\downarrow} + \gamma_{\uparrow})^2 + 4 \left((h_1 + h_2)^2 + \lambda^2\right)\right]}, \nonumber \\
O_{xy} &= \frac{2 \lambda (\gamma_{\uparrow} - \gamma_{\downarrow})}{(\gamma_{\downarrow} + \gamma_{\uparrow})^2 + 4 \left((h_1 + h_2)^2 + \lambda^2\right)} = O_{yx}, \nonumber \\
O_{yy} &= \frac{4 \lambda (\gamma_{\downarrow} - \gamma_{\uparrow})(h_1 + h_2)} {(\gamma_{\downarrow} + \gamma_{\uparrow}) \left[(\gamma_{\downarrow} + \gamma_{\uparrow})^2 + 4 \left((h_1 + h_2)^2 + \lambda^2\right)\right]}, \nonumber \\
O_{zz} &= \frac{(\gamma_{\downarrow} - \gamma_{\uparrow})^2 \left[(\gamma_{\downarrow} + \gamma_{\uparrow})^2 + 4 (h_1 + h_2)^2\right]}{4 (\gamma_{\downarrow} + \gamma_{\uparrow})^2 \left[(h_1 + h_2)^2 + \lambda^2\right] + (\gamma_{\downarrow} + \gamma_{\uparrow})^4}.
\end{align}
Note that for equal pump and loss rates, $\gamma_{\uparrow} = \gamma_{\downarrow}$, all observables vanish and the system remains in a trivial symmetric state. In the following, we consider the driven-dissipative regime with $\gamma_{\uparrow} = \gamma$ and $\gamma_{\downarrow} = 0$.

In order to resort to the linear analysis close to the steady-state solution, we linearize the equations of motion around the stationary solution 
$\vec{\nu}_S$. Specifically, we first rewrite the equations of motion \eqref{eom_all} in a compact form as

\begin{equation}
\frac{d \vec{X}}{d t} = \vec{F}(\vec{X})
\;\; , 
\label{Eom_X}
\end{equation}
\noindent
with 
\begin{align*}
\vec{X} = \{ & m_{x,1}, m_{x,2}, m_{y,1}, m_{y,2}, m_{z,1}, m_{z,2}, \\
             & O_{xx}, O_{yy}, O_{zz}, O_{xy}, O_{xz}, O_{yx}, O_{yz}, O_{zx}, O_{zy} \}
\end{align*}
\noindent
Next, we linearize Eqs.(\ref{Eom_X}) by introducing a small   perturbation around the steady-state solution $\vec{\nu}_S$, that is
\begin{equation}
\vec{X} = \vec{\nu}_S + \delta \vec{X} (t) \;\; .
\label{perturbation}
\end{equation}
\noindent
Substituting Eq.~\eqref{perturbation} into Eq.~\eqref{Eom_X} and keeping only terms linear in $\delta \vec{X}$, we obtain the corresponding equations of motion in the form 

\begin{equation}
\frac{d \delta \vec{X} (t) }{d t} = \mathcal{J}(\vec{\nu}_S)\, \delta \vec{X} (t) \;\;,
\label{linearization}
\end{equation}
\noindent
with   the Jacobian matrix   defined as \cite{Lukas_2024, Ott_2002, steven_2000}
\begin{equation}
\left. \mathcal{J}_{ij} \right|_{\vec{\nu}_S} \equiv \left. \frac{\partial F_i}{\partial X_j} \right|_{\vec{\nu}_S}
\:\: .
\label{jacob}
\end{equation}
\noindent
Note that, in contrast to Section~\ref{sec:DPT_A}, where we numerically compute the Jacobian in the density-matrix space, here we define and compute it directly from the equations of motion of the observables. This corresponds to a reduced dynamics in the observable space that includes only the physical collective modes, such as magnetizations and correlations \cite{Carmichael2007StatisticalMI}. Since the dissipative phase transition is a collective phenomenon, its signature must appear in the Jacobian defined in the observable space. As we evidence in the following, a change in the stability of a steady-state solution signals a dissipative phase transition. 
  
By direct inspection, we find that the nonzero elements of the Jacobian matrix are given by
\begin{align}
&2 \mathcal{J}_{1,1} = 2 \mathcal{J}_{2,2} = 2 \mathcal{J}_{3,3} = 2 \mathcal{J}_{4,4}  = \mathcal{J}_{5,5} = \mathcal{J}_{6,6} =
 \mathcal{J}_{7,7} \nonumber\\&= \mathcal{J}_{8,8}  =\frac{1}{2}\mathcal{J}_{9,9} = \mathcal{J}_{10,10} = \frac{2}{3} \mathcal{J}_{11,11} = \mathcal{J}_{12,12} = \frac{2}{3} \mathcal{J}_{13,13} \nonumber\\&= \frac{2}{3} \mathcal{J}_{14,14} = \frac{2}{3} \mathcal{J}_{15,15} = -\mathcal{J}_{14,2} = -\mathcal{J}_{13,3} = -\mathcal{J}_{15,4} \nonumber\\&= -\mathcal{J}_{11,1} = -\mathcal{J}_{9,5} = -\mathcal{J}_{9,6} = -g, \nonumber \\
&\mathcal{J}_{3,14} = \mathcal{J}_{4,11} = \mathcal{J}_{10,6} = \mathcal{J}_{12,5} = - \mathcal{J}_{5,12} = -\mathcal{J}_{6,10} \nonumber \\
&= -\mathcal{J}_{11,4} = -\mathcal{J}_{14,3} = 2 \lambda, \nonumber \\
&\mathcal{J}_{1,3} = \mathcal{J}_{7,12} = \mathcal{J}_{10,8} = \mathcal{J}_{11,13} = 2 h_1 = -\mathcal{J}_{8,10} = -\mathcal{J}_{12,7} \nonumber\\&= -\mathcal{J}_{13,11}, \nonumber \\
&\mathcal{J}_{2,4} = \mathcal{J}_{7,10} = \mathcal{J}_{12,8} = \mathcal{J}_{14,15} = 2 h_2 = -\mathcal{J}_{8,12} = -\mathcal{J}_{10,7} \nonumber\\&= -\mathcal{J}_{15,14}, \qquad \mathcal{J}_{11,2} = \mathcal{J}_{14,1} = - \frac{4 g \lambda}{d},\nonumber \\
&\mathcal{J}_{3,1} = 2 \left( -h_1 + \frac{g^2 + 4(h_1 + h_2)^2}{d} \right),\nonumber\\& \mathcal{J}_{4,2} = 2 \left( -h_2 + \frac{g^2 + 4(h_1 + h_2)^2}{d} \right), \nonumber \\
& \mathcal{J}_{13,1} = \mathcal{J}_{15,2} = 
\frac{2(g^2 + 4(h_1 + h_2)^2)}{d}, \nonumber \\
&\mathcal{J}_{13,2} = \mathcal{J}_{15,1} = \frac{8(h_1 + h_2) \lambda}{d}\:\:,
\end{align}
where $d = g^2 + 4 \left[(h_1 + h_2)^2 + \lambda^2 \right]$. The eigenvalues $ \mu_i$ of the Jacobian matrix evaluated at $\vec{\nu}_S$ determine the stability of the symmetric steady-state solution. If the real part of all eigenvalues is negative, $\Re(\mu_i) < 0$, the symmetric phase is stable. When one eigenvalue crosses zero, $\Re(\mu_{\mathrm{crit}}) = 0$, the symmetric phase loses stability and the broken-symmetry phase emerges. The point at which this occurs defines the critical point of the dissipative phase transition.

\begin{figure}
    \centering
    \includegraphics[width=0.8\linewidth]{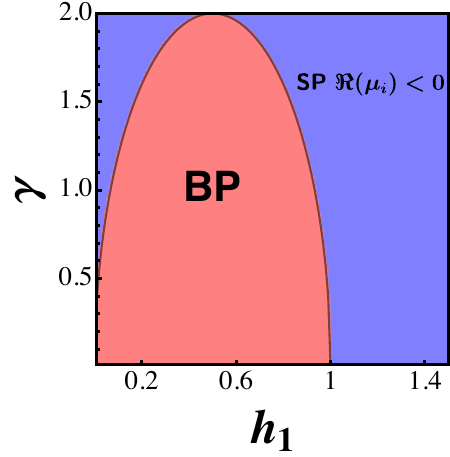}

    \vspace{0.2cm}

    \includegraphics[width=0.8\linewidth]{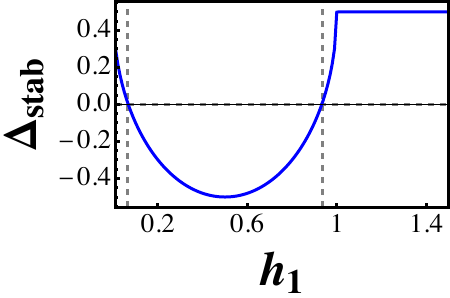}
           
    \caption{ Top: Phase diagram of the dissipative phase transition in the $h_1$–$\gamma$ plane. The red region indicates the symmetry-broken phase (BP), while the blue region corresponds to the symmetric phase (SP), where the real parts of all Jacobian eigenvalues are negative. The black line demarcates the phase boundary between the two phases. Bottom: Linear stability gap 
    $\Delta_{\mathrm{stab}}$ of the symmetric-phase solution as a function of $h_1$ for $\gamma = 1$. The parameters are fixed at
    $h_2 = 10$ and $\lambda = 0.01$. }
    \label{fig:DPT_Phase}
\end{figure}
\noindent
In Fig.~\ref{fig:DPT_Phase} (top panel), we present the dissipative phase diagram in the $h_1$–$\gamma$ plane, obtained for $h_2 = 10$ and $\lambda = 0.01$. The red region indicates the symmetry-broken phase, where the symmetric steady-state solution loses stability and a symmetry-broken state emerges. In contrast, the blue region corresponds to the symmetric phase, where all eigenvalues of the Jacobian have negative real parts. In the weak-dissipation limit, the system approaches the behavior expected for a closed quantum system, with the critical point located at $h_{1,c} \approx 1$. As $\gamma$ increases, the critical value $h_{1,c}$ required to sustain the symmetry-broken phase decreases, leading to a progressive reduction of the ordered region in the phase diagram.

For sufficiently strong dissipation ($\gamma > 0.5$), a distinct feature appears at low values of $h_1$. In this regime, dissipation dominates over the coherent dynamics, driving the system back into the symmetric phase (blue region) even for values of $h_1$ that would support the symmetry-broken phase at lower dissipation strengths. This behavior indicates that dissipation acts as an effective noise source that suppresses the ordered phase and stabilizes the symmetric phase. The appearance of this reentrant behavior reflects the competition between coherent interactions, external driving, and dissipation, which stabilizes the ordered phase only within a finite parameter window. A phase diagram similar to ours has also been reported in a long-range model in the regime where the interaction decay exponent $\alpha$ lies below the critical value for a one-dimensional system, $\alpha_c = 2$ \cite{Salatino:2026}.

In Fig.~\ref{fig:DPT_Phase}(bottom), we plot the linear stability gap $\Delta_{\mathrm{stab}} = - \max \Re(\mu)$ of the symmetric steady-state solution as a function of $h_1$ for a fixed value $\gamma = 1$. When $\Delta_{\mathrm{stab}} > 0$, all eigenvalues of the Jacobian have negative real parts and the symmetric solution is stable. When $\Delta_{\mathrm{stab}} < 0$, the real part of the leading eigenvalue becomes positive, indicating that the symmetric steady state becomes unstable and the system enters the broken symmetry phase. Thus, the region where $\Delta_{\mathrm{stab}} < 0$ corresponds to the broken phase. The critical points of the dissipative phase transition are identified by the points where $\Delta_{\mathrm{stab}}$ crosses zero. We find that the two critical points are located at approximately $h_1 \simeq 0.06$ and $h_1 \simeq 0.93$, for fixed values of $h_2 = 10$ and $\lambda = 0.01$. We also observe that the phase diagram depends sensitively on   $h_2$ and the interaction strength $\lambda$. Decreasing $h_2$ and that increasing $\lambda$ enhances the mean-field feedback, which stabilizes the ordered phase and enlarges the symmetry-broken region. However, for sufficiently large $\gamma$, dissipation again dominates and drives the system back into the symmetric phase. This behavior highlights the dissipative phase transition is governed by the competition between coherent mean-field feedback and dissipative damping.

Next, we analyze various steady-state observables and their dependence on   $h_1$ for a fixed value of  $\gamma$. The steady-state solutions for the observables are obtained numerically by setting the left-hand side of Eqs.~\eqref{eom_all} to zero and solving the resulting system of nonlinear algebraic equations. The stability of each solution is determined from the eigenvalues of the Jacobian matrix, and only the stable solutions are retained for different values of $h_1$, while keeping $\gamma = 1$, $h_2 = 10$, and $\lambda = 0.01$ fixed.
\begin{figure}
   \centering
    \includegraphics[width=0.85\linewidth]{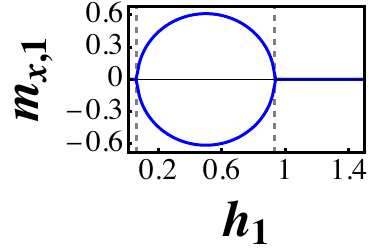}

    \vspace{0.2cm}

    \includegraphics[width=0.85\linewidth]{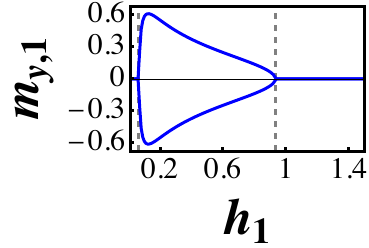}

    \vspace{0.2cm}

    \includegraphics[width=0.8\linewidth]{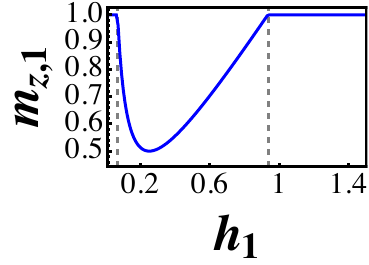}

    \caption{Dependence of the steady-state magnetizations $m_{x,1}$ (top), $m_{y,1}$ (middle), and $m_{z,1}$ (bottom) on the control parameter $h_1$, for fixed $\gamma = 1$, $h_2 = 10$, and $\lambda = 0.01$. The dashed black vertical lines indicate the critical points where the symmetric steady state loses stability and the symmetry-broken phase emerges.}
    \label{fig:mx1_my1_mz1}
\end{figure}
In Fig.~\ref{fig:mx1_my1_mz1}, we plot the steady-state magnetizations of model~1 as a function of $h_1$. For sufficiently strong dissipation ($\gamma = 1$), $m_{x,1}$ and $m_{y,1}$ vanish at small $h_1$, indicating a symmetric  phase, while the longitudinal magnetization $m_{z,1}$ remains finite. Upon increasing $h_1$, a symmetry-broken phase emerges in the interval $h_1 \in (0.07,\,0.93)$, where $m_{x,1}$ acquires one of two finite values corresponding to two symmetry-related steady states. Due to the finite inter-model coupling $\lambda$, the transverse magnetization in the model~2 is induced by the model~1, leading to a small but finite value $m_{x,2} \simeq (\lambda/h_2)\, m_{x,1}$ in the broken phase. Upon further increasing $h_1$, at a second critical point the symmetric steady state regains stability and the symmetry-broken phase disappear. The nonzero values of $m_{x,1}$ and $m_{y,1}$ between the two critical points form the bifurcation diagram of the symmetry-broken phase. The transition is associated with a pitchfork bifurcation of the steady-state solutions, where the symmetric steady state loses stability and two symmetry-broken steady states emerge continuously, indicating a continuous dissipative phase transition. This results in a reentrant symmetry-broken phase bounded by two critical points. We have verified that the reentrant behavior persists over a finite range of parameters, indicating that it is not a fine-tuned effect. This non-monotonic behavior indicates that the transverse field plays a dual role in the dissipative dynamics. While a finite field is required to generate the coherences necessary for stabilizing the ordered phase, an excessively strong field ultimately suppresses the $x$-magnetization by polarizing the spins along the $z$-direction. 
    
From the equations of motion in Eqs.~\eqref{eom_all}, one can see that the transverse field $h_1$ mixes the magnetization components in all three directions $m_{\alpha,1}$ with $\alpha = x,y,z$. This mixing induces coherences between spin components, which are necessary to convert interaction-generated correlations into a finite order parameter $m_{x,1}$. Since the interaction term alone does not mix the $\sigma^z$ and $\sigma^x$ sectors, a finite transverse field is required to generate the coherences that allow the interaction to stabilize a nonzero magnetization. For sufficiently strong dissipation $\gamma$, the system cannot sustain a finite $m_{x,1}$ at low values of $h_1$, as the order parameter is suppressed by dissipation. As a result, the symmetry-broken phase appears only above a finite threshold value of $h_1$. On the other hand, for very large $h_1$, the strong field polarizes the spins along the $z$ direction, thereby suppressing the order parameter $m_{x,1}$. This competition leads to a reentrant phase diagram, where the system exhibits a symmetry-broken phase bounded by two continuous dissipative phase transitions.

In the equilibrium version of the model, no reentrant phase is present. Remarkably, the inclusion of dissipation leads to the emergence of such a reentrant phase, highlighting a qualitatively new effect induced by nonequilibrium dynamics. This demonstrates that dissipation does not merely suppress ordering, but can fundamentally reshape the phase structure by stabilizing ordered phases only within a finite window of control parameters.
 
\section{Conclusion}
\label{concl}
We have studied phase transitions in a model of open quantum systems coupled to an external environment. The dissipative dynamics is described within the Lindblad master equation framework, incorporating both coherent unitary evolution and incoherent dissipative processes. The coherent part is governed by a time-dependent, self-consistent mean-field Hamiltonian, while the dissipative contribution is constructed from pertinently chosen jump operators.

 We have first analyzed the post-quench relaxation dynamics when the jump operators are defined in the instantaneous eigenbasis of the mean-field Hamiltonian and satisfy detailed balance. In this case, the system exhibits conventional relaxation for parametric quenches of the Hamiltonian, whereas temperature quenches give rise to dynamical phase transitions characterized by nonanalytic behavior in time. Increasing the dissipation strength progressively smears out these nonanalytic features. 

 We then turn to the steady-state properties and the associated dissipative phase transitions. When the jump operators are defined in the instantaneous eigenbasis of the mean-field Hamiltonian, the dissipative dynamics becomes effectively self-consistent and nonlinear. Under detailed balance, the system evolves toward an asymptotic steady state that closely resembles a thermal state in the eigenbasis of the Hamiltonian, and the resulting dissipative phase transition mimics the equilibrium one of the corresponding closed system. Nevertheless, linear stability analysis of the thermal dissipator indicates the absence of true critical slowing down at the critical point. Instead, the relaxation dynamics exhibits a pronounced but nondivergent slowing.
 
In contrast, when the dissipative dynamics is generated by local spin raising and lowering operators, the system is driven into a genuinely nonequilibrium steady state. In this case, while the dissipative phase transition coincides with the equilibrium critical point at weak system–bath coupling, increasing the dissipation strength leads to a significantly richer phase diagram. In particular, we observe a reentrant phase transition driven entirely by dissipative effects. Such behavior highlights the nontrivial role of dissipation in both stabilizing and destabilizing ordered phases. From the behavior of local observables, we find that the transition in the local dissipator case is associated with a pitchfork bifurcation of the steady-state solutions, whereas no such bifurcation occurs when dissipation enforces detailed balance.

 Although our analysis is based on a minimal model, the underlying mechanisms are expected to be generic. Overall, our results demonstrate how the nature of dissipation fundamentally controls both the dynamical and steady-state critical behavior in open quantum systems. 

From a broader perspective, our work shows that linear stability analysis provides a natural framework to characterize critical behavior in nonlinear open quantum systems, where a global Liouvillian description is not available. An interesting direction for future research would be to go beyond the mean-field approximation and investigate the role of fluctuations and spatial correlations in finite-dimensional systems. These directions may help clarify how the interplay between dissipation, and interactions shapes nonequilibrium critical phenomena.

\section{DATA AVAILABILITY}
The data underlying the figures in this work are available at Zenodo \cite{zenodo}.

\begin{acknowledgments}
A. N. acknowledges funding by the Deutsche Forschungsgemeinschaft (DFG, German Research Foundation) under Projektnummer 277101999 - TRR 183 (project  B02), under Projektnummer EG 96/14-1, and under Germany's Excellence Strategy - Cluster of Excellence Matter and Light for Quantum Computing (ML4Q) EXC 2004/2 - 390534769.
\end{acknowledgments}

%%%%%%%%%%%%%%%%%%%%%%%%%%%%%%%%%%%%%%%%%%%%%%%%%%%%%%%%%%%%%%%%%%%%%%%%%%%%%%%%%%%%%%%%%%%%%
% Appendix
%%%%%%%%%%%%%%%%%%%%%%%%%%%%%%%%%%%%%%%%%%%%%%%%%%%%%%%%%%%%%%%%%%%%%%%%%%%%%%%%%%%%%%%%%%%%%
\appendix
\section{Initial thermal state preparation}
\label{Initial_thermal}
To prepare an initial thermal state, we begin with an ansatz density matrix taken as a product state of the two subsystems,
\begin{align}
\hat{\rho}_i = \rho_1 \otimes \rho_2 ,
\end{align}
where the subsystem density matrices are chosen as
\begin{align}
\rho_{1} =\frac{1}{2}\begin{pmatrix} 
1+\tanh(\beta h_1) & \epsilon \\ 
\epsilon & 1-\tanh(\beta h_1) 
\end{pmatrix},
\end{align}
and
\begin{align}
\rho_{2} =\frac{1}{2}\begin{pmatrix} 
1+\tanh(\beta h_2) & \epsilon \\
\epsilon & 1-\tanh(\beta h_2)
\end{pmatrix}.
\end{align}
Here, $\beta = 1/T> 1.$ denotes the inverse temperature, and $\epsilon$ is a small symmetry-breaking seed introduced to allow spontaneous symmetry breaking during the subsequent evolution. The smallness of $\epsilon$ ensures that the initial magnetization in the $x$-direction is nearly zero.

We then choose the initial parameters $h_{1,i}$ and $T_i$, and numerically integrate the self-consistent mean-field Lindblad equations [Eqs.~\eqref{eq:nonlinear_lindblad}, \eqref{eq:Main_MF_H}, and \eqref{eq:m1}] for sufficiently long times until the density matrix converges to a steady state. Here, we set $\gamma^{\Downarrow}_{\mathrm{high}} = \gamma^{\Downarrow}_{\mathrm{low}}$ in order to minimize the preparation time. Since the jump operators are defined in the instantaneous eigenbasis of the Hamiltonian $H$, with transition rates satisfying detailed balance, the long-time steady state takes the Boltzmann form
\begin{align}
\rho_{\rm ss} \equiv \rho_i = \frac{e^{-H/T_i}}{Z},
\end{align}
where
\begin{align}
Z = \mathrm{Tr}\!\left(e^{-H/T_i}\right)
\end{align}
is the partition function.

Importantly, this thermal state is obtained via dissipative relaxation and is subsequently used as an initial condition; it does not imply that the system remains coupled to an equilibrium bath during the post-quench dynamics. The state $\rho_i$ is therefore interpreted as a thermal state at temperature $T_i$ and is used as the initial condition for the quench protocol.

Depending on the chosen values of $h_{1,i}$ and $T_i$, the prepared state may lie either in the ordered phase, characterized by a finite magnetization in the $x$-direction, or in the disordered phase where the transverse  magnetization vanishes. In the ordered phase, because we work in the regime $h_2 \gg h_1 \gg \lambda$, a finite magnetization $m_{x,1}$ in the first subsystem induces a small but nonzero transverse magnetization $m_{x,2}$ in the second subsystem through the weak inter-model coupling, even when $h_2 > h_{2,c}(T)$.

At time $t=0$, the system parameters and dissipative parameters (e.g., bath temperature) are changed, and the density matrix evolves under the post-quench Lindblad master equation. In this way, we study driven-dissipative relaxation dynamics starting from a thermal initial state.

 \bibliography{refs}
\end{document}